*Sequence Alignment*

# Shouji: A Fast and Efficient Pre-Alignment Filter for Sequence Alignment

Mohammed Alser[1,2,3,*], Hasan Hassan[1], Akash Kumar[2],
Onur Mutlu[1,3,*], and Can Alkan[3,*]
[1]Computer Science Department, ETH Zürich, 8092 Zürich, Switzerland,
[2]Institute for Computer Engineering, CfAED, Technische Universität Dresden, Germany,
[3]Computer Engineering Department, Bilkent University, 06800 Bilkent, Ankara, Turkey

*To whom correspondence should be addressed.

## Abstract

**Motivation:** The ability to generate massive amounts of sequencing data continues to overwhelm the processing capability of existing algorithms and compute infrastructures. In this work, we explore the use of hardware/software co-design and hardware acceleration to significantly reduce the execution time of short sequence alignment, a crucial step in analyzing sequenced genomes. We introduce Shouji, a highly-parallel and accurate pre-alignment filter that remarkably reduces the need for computationally-costly dynamic programming algorithms. The first key idea of our proposed pre-alignment filter is to provide high filtering accuracy by correctly detecting all common subsequences shared between two given sequences. The second key idea is to design a hardware accelerator that adopts modern FPGA (Field-Programmable Gate Array) architectures to further boost the performance of our algorithm.

**Results:** Shouji significantly improves the accuracy of pre-alignment filtering by up to two orders of magnitude compared to the state-of-the-art pre-alignment filters, GateKeeper and SHD. Our FPGA-based accelerator is up to three orders of magnitude faster than the equivalent CPU implementation of Shouji. Using a single FPGA chip, we benchmark the benefits of integrating Shouji with five state-of-the-art sequence aligners, designed for different computing platforms. The addition of Shouji as a pre-alignment step reduces the execution time of the five state-of-the-art sequence aligners by up to 18.8x. Shouji can be adapted for any bioinformatics pipeline that performs sequence alignment for verification. Unlike most existing methods that aim to accelerate sequence alignment, Shouji does *not* sacrifice any of the aligner capabilities, as it does *not* modify or replace the alignment step.

**Availability:** https://github.com/CMU-SAFARI/Shouji

**Contact:** mohammed.alser@inf.ethz.ch, onur.mutlu@inf.ethz.ch, calkan@cs.bilkent.edu.tr

**Supplementary information:** Supplementary data are available at *Bioinformatics* online.

## 1 Introduction

One of the most fundamental computational steps in most bioinformatics analyses is the detection of the differences/similarities between two genomic sequences. *Edit distance* and *pairwise alignment* are two approaches to achieve this step, formulated as *approximate string matching* (Navarro, 2001). Edit distance approach is a measure of how much two sequences differ. It calculates the minimum number of edits needed to convert a sequence into the other. The higher the edit distance, the more different the sequences from one another. Commonly-allowed edit operations include deletion, insertion, and substitution of characters in one or both sequences. Pairwise alignment is a measure of how much the sequences are alike. It calculates the alignment that is an ordered list of characters representing possible edit operations and matches required to change one of the two given sequences into the other. As any two sequences can have several different arrangements of the edit operations and matches (and hence different alignments), the alignment algorithm usually involves a backtracking step. This step finds the alignment that has the highest *alignment score* (called *optimal* alignment). The alignment score is the sum of the scores of all edits and matches along the alignment implied by a user-defined *scoring function*. The edit distance and pairwise alignment approaches are *non-additive measures* (Calude et al., 2002). This means that if we divide the sequence pair into two consecutive subsequence pairs, the edit distance of the entire sequence pair is not necessarily equivalent to the sum of the edit distances of the shorter pairs. Instead, we need to examine all possible *prefixes* of the two input sequences and keep track of the pairs of prefixes that provide an optimal solution. Enumerating all possible prefixes is necessary for tolerating edits that result from both sequencing errors (Fox et al., 2014) and genetic variations (McKernan et al., 2009). Therefore, the edit distance and pairwise alignment approaches are typically implemented as dynamic programming algorithms to avoid re-examining the same prefixes many times. These implementations, such as Levenshtein distance (Levenshtein, 1966), Smith-Waterman (Smith and Waterman, 1981), and Needleman-Wunsch (Needleman and Wunsch, 1970), are inefficient as they have quadratic time and space complexity (i.e., $O(m^2)$ for a sequence length of $m$). Many attempts were made to boost the performance of existing sequence aligners. Despite more than three decades of attempts, the fastest known edit distance algorithm (Masek and Paterson, 1980) has a running time of $O(m^2/\log^2 m)$ for sequences of length $m$, which is still nearly quadratic (Backurs and Indyk, 2017). Therefore, more recent works tend to follow one of two key new directions to boost the performance of sequence alignment and edit distance implementations: (1) Accelerating the dynamic programming algorithms using hardware accelerators. (2) Developing *filtering heuristics* that reduce the need for the dynamic programming algorithms, given an edit distance threshold.

**Hardware accelerators are becoming increasingly popular for speeding up the computationally-expensive alignment and edit distance algorithms (Al Kawam et al., 2017; Aluru and Jammula, 2014; Ng et al., 2017; Sandes et al., 2016).** Hardware accelerators include multi-core and SIMD (single instruction multiple data) capable central processing units (CPUs), graphics processing units (GPUs), and field-programmable gate arrays (FPGAs). The classical dynamic programming algorithms are typically accelerated by computing *only* the necessary regions (i.e., diagonal vectors) of the dynamic programming matrix rather than the entire matrix, as proposed in Ukkonen's banded algorithm (Ukkonen, 1985). The number of the diagonal bands required for computing the dynamic programming matrix is $2E+1$, where $E$ is a user-defined edit distance threshold. The banded algorithm is still beneficial even with its recent sequential implementations as in Edlib (Šošić and Šikić, 2017). The Edlib algorithm is implemented in C for standard CPUs and it calculates the banded Levenshtein distance. Parasail (Daily, 2016) exploits both Ukkonen's banded algorithm and SIMD-capable CPUs to compute a *banded alignment* for a sequence pair with a user-defined scoring function. SIMD instructions offer significant parallelism to the matrix computation by executing the same vector operation on *multiple operands* at once. The multi-core architecture of CPUs and GPUs provides the ability to compute alignments of *many sequence pairs* independently and concurrently (Georganas et al., 2015; Liu and Schmidt, 2015). GSWABE (Liu and Schmidt, 2015) exploits GPUs (Tesla K40) for highly-parallel computation of global alignment with a user-defined scoring function. CUDASW++ 3.0 (Liu et al., 2013) exploits the SIMD capability of both CPUs and GPUs (GTX690) to accelerate the computation of the Smith-Waterman algorithm with a user-defined scoring function. CUDASW++ 3.0 provides only the optimal score, not the optimal alignment (i.e., no backtracking step). Other designs, for instance FPGASW (Fei et al., 2018), exploit the very large number of hardware execution units in FPGAs (Xilinx VC707) to form a linear systolic array (Kung, 1982). Each execution unit in the systolic array is responsible for computing the value of a single entry of the dynamic programming matrix. The systolic array computes a single vector of the matrix at a time. The data dependencies between the entries restrict the systolic array to computing the vectors sequentially (e.g., top-to-bottom, left-to-right, or in an anti-diagonal manner). FPGA accelerators seem to yield the highest performance gain compared to the other hardware accelerators (Banerjee et al., 2018; Chen et al., 2016; Fei et al., 2018; Waidyasooriya and Hariyama, 2015). However, many of these efforts either *simplify* the scoring function, or only take into account accelerating the computation of the dynamic programming matrix *without* providing the optimal alignment as in (Chen et al., 2014; Liu et al., 2013; Nishimura et al., 2017). Different and more sophisticated scoring functions are typically needed to better quantify the similarity between two sequences (Henikoff and Henikoff, 1992; Wang et al., 2011). The backtracking step required for the optimal alignment computation involves unpredictable and irregular memory access patterns, which poses a difficult challenge for efficient hardware implementation.

**Pre-alignment filtering heuristics aim to quickly eliminate some of the dissimilar sequences *before* using the computationally-expensive optimal alignment algorithms.** There are a few existing filtering techniques such as the Adjacency Filter (Xin et al., 2013), which is implemented for standard CPUs as part of FastHASH (Xin et al., 2013). SHD (Xin et al., 2015) is a SIMD-friendly bit-vector filter that provides higher filtering accuracy compared to the Adjacency Filter. GRIM-Filter (Kim et al., 2018) exploits the high memory bandwidth and the logic layer of 3D-stacked memory to perform highly-parallel filtering in the DRAM chip itself. GateKeeper (Alser et al., 2017) is designed to utilize the large amounts of parallelism offered by FPGA architectures. MAGNET (Alser et al., July 2017) shows a low number of falsely-accepted sequence pairs but its current implementation is much slower than that of SHD or GateKeeper. GateKeeper (Alser et al., 2017) provides a high filtering speed but suffers from relatively high number of falsely-accepted sequence pairs.

**Our goal** in this work is to significantly reduce the time spent on calculating the *optimal alignment* of short sequences and maintain high filtering accuracy. To this end, we introduce Shouji[1], a new, fast, and very accurate pre-alignment filter. Shouji is based on two key ideas: (1) A new filtering algorithm that remarkably reduces the need for computationally-expensive banded optimal alignment by *rapidly excluding dissimilar sequences from the optimal alignment calculation*. (2) Judicious use of the parallelism-friendly architecture of modern FPGAs to greatly speed up this new filtering algorithm.

The contributions of this paper are as follows:

- We introduce Shouji, a highly-parallel and highly-accurate pre-alignment filter, which uses a *sliding search window approach* to quickly identify dissimilar sequences *without* the need for computationally-expensive alignment algorithms. We overcome the implementation limitations of MAGNET (Alser et al., July 2017). We build two hardware accelerator designs that adopt modern FPGA architectures to boost the performance of both Shouji and MAGNET.
- We provide a comprehensive analysis of the run time and space complexity of Shouji and MAGNET algorithms. Shouji and MAGNET are asymptomatically *inexpensive* and run in linear time with respect to the sequence length and the edit distance threshold.

[1] Named after a traditional Japanese door that is designed to slide open http://www.aisf.or.jp/~jaanus/deta/s/shouji.htm.

- We demonstrate that Shouji and MAGNET significantly improve the accuracy of pre-alignment filtering by up to two and four orders of magnitude, respectively, compared to GateKeeper and SHD.
- We demonstrate that our FPGA implementations of Shouji and MAGNET are two to three orders of magnitude faster than their CPU implementations. We demonstrate that integrating Shouji with five state-of-the-art aligners reduces the execution time of the sequence aligner by up to 18.8x.

# 2 METHODS

## 2.1 Overview

Our goal is to quickly reject dissimilar sequences with high accuracy such that we reduce the need for the computationally-costly alignment step. To this end, we propose the Shouji algorithm to achieve highly-accurate filtering. Then, we accelerate Shouji by taking advantage of the parallelism of FPGAs to achieve fast filtering operations. The key filtering strategy of Shouji is inspired by the ***pigeonhole principle***, which states that if $E$ items are distributed into $E$+1 boxes, then one or more boxes would remain empty. In the context of pre-alignment filtering, this principle provides the following key observation: if two sequences differ by $E$ edits, then the two sequences should share *at least* a single common subsequence (i.e., free of edits) and *at most* $E$+1 non-overlapping common subsequences, where $E$ is the edit distance threshold. With the existence of at most $E$ edits, the total length of these non-overlapping common subsequences should *not* be less than $m$-$E$, where $m$ is the sequence length. Shouji employs the pigeonhole principle to decide whether or not two sequences are potentially similar. Shouji finds all the non-overlapping subsequences that exist in both sequences. If the total length of these common subsequences is less than $m$-$E$, then there exist more edits than the allowed edit distance threshold, and hence Shouji rejects the two given sequences. Otherwise, Shouji accepts the two sequences. Next, we discuss the details of Shouji.

## 2.2 Shouji Pre-alignment Filter

Shouji identifies the dissimilar sequences, without calculating the optimal alignment, in three main steps. (1) The first step is to construct what we call a *neighborhood map* that visualizes the pairwise matches and mismatches between two sequences given an edit distance threshold of $E$ characters. (2) The second step is to find all the non-overlapping common subsequences in the neighborhood map using a sliding search window approach. (3) The last step is to accept or reject the given sequence pairs based on the length of the found matches. If the length of the found matches is small, then Shouji rejects the input sequence pair.

### 2.2.1 Building the Neighborhood Map

The neighborhood map, $N$, is a binary $m$ by $m$ matrix, where $m$ is the sequence length. Given a text sequence $T[1...m]$, a pattern sequence $P[1...m]$, and an edit distance threshold $E$, the neighborhood map represents the comparison result of the $i^{th}$ character of $P$ with the $j^{th}$ character of $T$, where $i$ and $j$ satisfy $1 \le i \le m$ and $i\text{-}E \le j \le i\text{+}E$. The entry $N[i, j]$ of the neighborhood map can be calculated as follows:

$$N[i,j] = \begin{cases} 0, & \text{if } P[i] = T[j] \\ 1, & \text{if } P[i] \neq T[j] \end{cases} \tag{1}$$

We present in Fig. 1 an example of a neighborhood map for two sequences, where a pattern $P$ differs from a text $T$ by three edits.

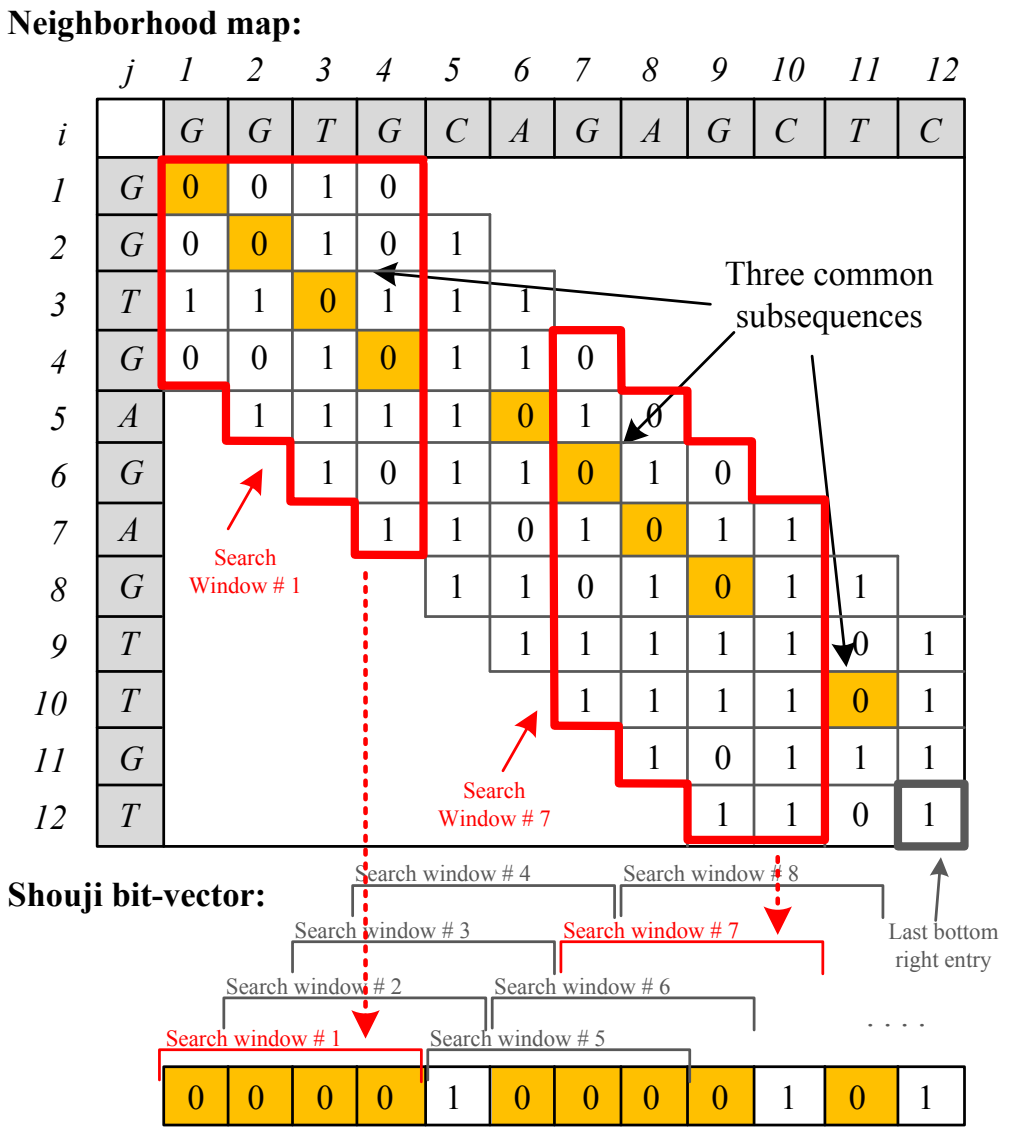

**Fig. 1: Neighborhood map (*N*) and the Shouji bit-vector, for text *T* = GGTGCAGAGCTC, and pattern *P* = GGTGAGAGTTGT for *E*=3. The three common subsequences (i.e., GGTG, AGAG, and T) are highlighted in yellow. We use a search window of size 4 columns (two examples of which are highlighted in red) with a step size of a single column. Shouji searches diagonally within each search window for the 4-bit vector that has the largest number of zeros. Once found, Shouji examines if the found 4-bit vector maximizes the number of zeros at the corresponding location of the 4-bit vector in the Shouji bit-vector. If so, then Shouji stores this 4-bit vector in the Shouji bit-vector at its corresponding location.**

The entry $N[i, j]$ is set to zero if the $i^{th}$ character of the pattern matches the $j^{th}$ character of the text. Otherwise, it is set to one. The way we build our neighborhood map ensures that computing each of its entries is independent of every other, and thus the entire map can be computed all at once in a parallel fashion. Hence, our neighborhood map is well suited for highly-parallel computing platforms (Alser et al., 2017; Seshadri et al., 2017). Note that in sequence alignment algorithms, computing each entry of the dynamic programming matrix depends on the values of the immediate left, upper left, and upper entries of its own. Different from "dot plot" or "dot matrix" (visual representation of the similarities between two closely similar genomic sequences) that is used in FASTA/FASTP (Lipman and Pearson, 1985), our neighborhood map computes *only* necessary diagonals near the main diagonal of the matrix (e.g., seven diagonals shown in Fig. 1).

### 2.2.2 Identifying the Diagonally-Consecutive Matches

The key goal of this step is to accurately find all the non-overlapping common subsequences shared between a pair of sequences. The accuracy of finding these subsequences is crucial for the overall filtering accuracy, as the filtering decision is made solely based on total subsequence length. With the existence of $E$ edits, there are *at most* $E$+1 non-overlapping common subsequences (based on the pigeonhole principle) shared between a pair of sequences. Each non-overlapping common subsequence is represented as a streak of diagonally-consecutive zeros in the neighborhood map (as highlighted in yellow in Fig. 1). These streaks of diagonally-consecutive zeros are distributed along the diagonals of the neighborhood map without any prior information about their length or number. One way of finding these common subsequences is to use a brute-force approach, which examines all the streaks of diagonally-consecutive zeros that start at the first column and selects the streak that has the largest number of zeros as the first common subsequences. It then iterates over the remaining part of the neighborhood map to find the other common subsequences. However, this brute-force approach is infeasible for highly-optimized hardware implementation as the search space is unknown at design time. Shouji overcomes this issue by dividing the neighborhood map into equal-size parts. We call each part a *search window*. *Limiting the size of the search space* from the entire neighborhood map to a search window has three key benefits. (1) It helps to provide a scalable architecture that can be implemented for any sequence length and edit distance threshold. (2) Downsizing the search space into a reasonably small sub-matrix with a known dimension at design time limits the number of all possible permutations of each bit-vector to $2^n$, where $n$ is the search window size. This reduces the size of the look-up tables (LUTs) required for an FPGA implementation and simplifies the overall design. (3) Each search window is considered as a smaller sub-problem that can be solved independently and rapidly with high parallelism. Shouji uses a search window of 4 columns wide, as we illustrate in Fig. 1. We need $m$ search windows for processing two sequences, each of which is of length $m$ characters. Each search window overlaps with its next neighboring search window by 3 columns. This ensures covering the entire neighborhood map and finding all the common subsequences regardless of their starting location. We select the width of each search window to be 4 columns to guarantee finding the shortest possible common subsequence, which is a single match located between two mismatches (i.e., '101'). However, we observe that the bit pattern '101' is *not* always necessarily a part of the correct alignment (or the common subsequences). For example, the bit pattern '101' exists once as a part of the correct alignment in Fig.1, but it also appears five times in other different locations that are *not* included in the correct alignment. To improve the accuracy of finding the diagonally-consecutive matches, we increase the length of the diagonal vector to be examined to four bits. We also experimentally evaluate different search window sizes in Supplementary Materials, Section 6.1. We find that a search window size of 4 columns provides the highest filtering accuracy without falsely-rejecting similar sequences.

Shouji finds the diagonally-consecutive matches that are part of the common subsequences in the neighborhood map in two main steps. **Step 1:** For each search window, Shouji finds a 4-bit diagonal vector that has the largest number of zeros. Shouji greedily considers this vector as a part of the common subsequence as it has the least possible number of edits (i.e., 1's). Finding always the maximum number of matches is necessary to avoid overestimating the actual number of edits and eventually preserving all similar sequences. Shouji achieves this step by comparing the 4 bits of each of the $2E+1$ diagonal vectors within a search window and selects the 4-bit vector that has the largest number of zeros. In the case where two 4-bit subsequences have the same number of zeros, Shouji breaks the ties by selecting the first one that has a leading zero. Then, Shouji slides the search window by a single column (i.e., step size = 1 column) towards the last bottom right entry of the neighborhood map and repeats the previous computations. Thus, Shouji performs "Step 1" $m$ times using $m$ search windows, where $m$ is the sequence length. **Step 2:** The last step is to gather the results found for each search window (i.e., 4-bit vector that has the largest number of zeros) and construct back all the diagonally-consecutive matches. For this purpose, Shouji maintains a *Shouji bit-vector* of length $m$ that stores all the zeros found in the neighborhood map as we illustrate in Fig. 1. For each sliding search window, Shouji examines if the selected 4-bit vector maximizes the number of zeros in the Shouji bit-vector at the same corresponding location. If so, Shouji stores the selected 4-bit vector in the Shouji bit-vector at the same corresponding location. This is necessary to avoid overestimating the number of edits between two given sequences. The common subsequences are represented as streaks of consecutive zeros in the Shouji bit-vector.

#### 2.2.3 Filtering out Dissimilar Sequences

The last step of Shouji is to calculate the total number of edits (i.e., ones) in the Shouji bit-vector. Shouji examines if the total number of ones in the Shouji bit-vector is greater than $E$. If so, Shouji excludes the two sequences from the optimal alignment calculation. Otherwise, Shouji considers the two sequences similar within the allowed edit distance threshold and allows their optimal alignment to be computed using optimal alignment algorithms. The Shouji bit-vector represents the differences between two sequences along the entire length of the sequence, $m$. However, Shouji is not limited to end-to-end edit distance calculation. Shouji is also able to provide edit distance calculation in local and glocal (semi-global) fashion. For example, achieving local edit distance calculation requires ignoring the ones that are located at the two ends of the Shouji bit-vector. We present an example of local edit distance between two sequences of different length in Supplementary Materials, Section 8. Achieving glocal edit distance calculation requires excluding the ones that are located at one of the two ends of the Shouji bit-vector from the total count of the ones in the Shouji bit-vector. This is important for correct pre-alignment filtering for global, local, and glocal alignment algorithms. We provide the pseudo-code of Shouji and discuss its computational complexity in Supplementary Materials, Section 6.2. We also present two examples of applying the Shouji filtering algorithm in Supplementary Materials, Section 8.

## 2.3 Accelerator Architecture

Our second aim is to substantially accelerate Shouji, by leveraging the parallelism of FPGAs. In this section, we present our hardware accelerator that is designed to exploit the large amounts of parallelism offered by modern FPGA architectures (Aluru and Jammula, 2014; Herbordt et al., 2007; Trimberger, 2015). We then outline the implementation of Shouji to be used in our accelerator design. Fig. 2 shows the hardware architecture of the accelerator. It contains a user-configurable number of filtering units. Each filtering unit provides pre-alignment filtering independently from other units. The workflow of the accelerator starts with transmitting the sequence pair to the FPGA through the fastest communication medium available on the FPGA board (i.e., PCIe). The sequence controller manages and provides the necessary input signals for each filtering unit in the accelerator. Each filtering unit requires two sequences of the same length and an edit distance threshold. The result controller gathers the output result (i.e., a single bit of value '1' for similar sequences and '0' for dissimilar sequences) of each filtering unit and transmits them back to the host side in the same order as their sequences are transmitted to the FPGAs.

The host-FPGA communication is achieved using RIFFA 2.2 (Jacobsen et al., 2015). To make the best use of the available resources in the FPGA chip, our algorithm utilizes the operations that are easily supported on an FPGA, such as bitwise operations, bit shifts, and bit count. To build the neighborhood map on the FPGA, we use the observation that the main diagonal can be implemented using a bitwise XOR operation between the two given sequences. The upper $E$ diagonals can be implemented by gradually shifting the pattern ($P$) to the right-hand direction and then performing bitwise XOR with the text ($T$). This allows each character of $P$ to be compared with the right-hand neighbor characters (up to $E$ characters) of its corresponding character of $T$. The lower $E$ diagonals can be implemented in a way similar to the upper $E$ diagonals, but here the shift oper-ation is performed in the left-hand direction. This ensures that each character of *P* is compared with the left-hand neighbor characters (up to *E* characters) of its corresponding character of *T*.

We also build an efficient hardware architecture for each search window of the Shouji algorithm. It quickly finds the number of zeros in each 4-bit vector using a hardware look-up table that stores the 16 possible permutations of a 4-bit vector along with the number of zeros for each permutation. We present the block diagram of the search window architecture in Supplementary Materials, Section 6.3. Our hardware implementation of the Shouji filtering unit is independent of the specific FPGA-platform as it does not rely on any vendor-specific computing elements (e.g., intellectual property cores). However, each FPGA board has different resources and hardware capabilities that can directly or indirectly affect the performance and the data throughput of the design. The maximum data throughput of the design and the available FPGA resources determine the number of filtering units in the accelerator. Thus, if, for example, the memory bandwidth is saturated, then increasing the number of filtering units would not improve performance.

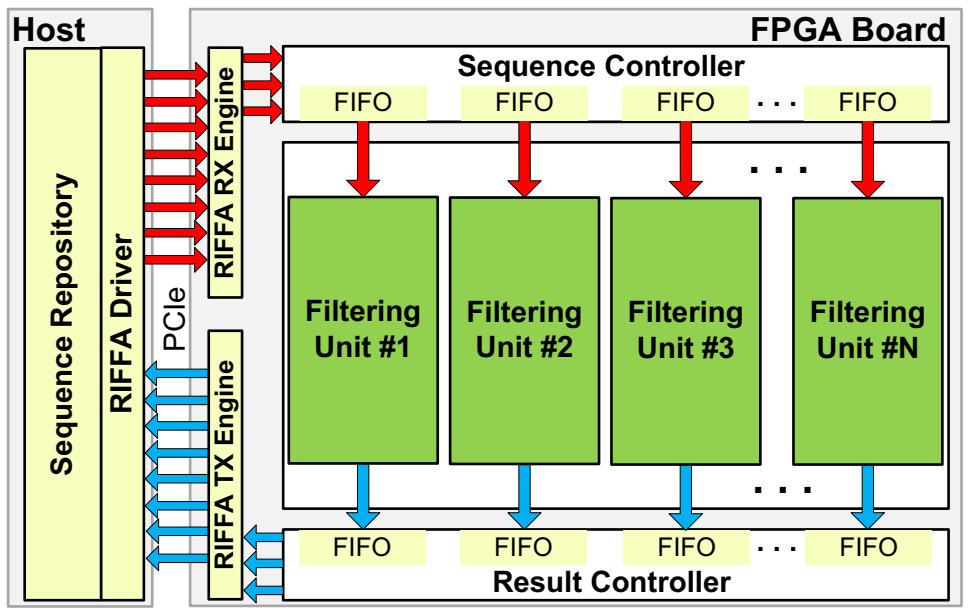

**Fig. 2: Overview of our hardware accelerator architecture. The filtering units can be replicated as many times as possible based on the resources available on the FPGA.**

## 3 RESULTS

In this section, we evaluate (1) the filtering accuracy, (2) the FPGA resource utilization, (3) the execution time of Shouji, our hardware implementation of MAGNET (Alser et al., July 2017), GateKeeper (Alser et al., 2017), and SHD (Xin et al., 2015), (4) the benefits of the pre-alignment filters together with state-of-the-art aligners, and (5) the benefits of Shouji together with state-of-the-art read mappers. As we mention in Section 1, MAGNET leads to a small number of falsely-accepted sequence pairs but suffers from poor performance. We comprehensively explore this algorithm and provide an efficient and fast hardware implementation of MAGNET in Supplementary Materials, Section 7. We run all experiments using a 3.6 GHz Intel i7-3820 CPU with 8 GB RAM. We use a Xilinx Virtex 7 VC709 board (Xilinx, 2014) to implement our accelerator architecture (for both Shouji and MAGNET). We build the FPGA design using Vivado 2015.4 in synthesizable Verilog.

### 3.1 Dataset Description

Our experimental evaluation uses 12 different real datasets. Each dataset contains 30 million real sequence pairs. We obtain three different read sets (ERR240727_1, SRR826460_1, and SRR826471_1) of the whole human genome that include three different read lengths (100 bp, 150 bp, and 250 bp). We download these three read sets from EMBL-ENA (www.ebi.ac.uk/ena). We map each read set to the human reference genome (GRCh37) using the mrFAST (Alkan et al., 2009) mapper. We obtain the human reference genome from the 1000 Genomes Project (Consortium, 2012). For each read set, we use four different maximum numbers of allowed edits using the *-e* parameter of mrFAST to generate four real datasets. Each dataset contains the sequence pairs that are generated by the mrFAST mapper before the read alignment step. This enables us to measure the effectiveness of the filters using both aligned and unaligned sequences over a wide range of edit distance thresholds. We summarize the details of these 12 datasets in Supplementary Materials, Section 9. For the reader's convenience, when referring to these datasets, we number them from 1 to 12 (e.g., set_1 to set_12). We use Edlib (Šošić and Šikić, 2017) to generate the ground truth edit distance value for each sequence pair.

### 3.2 Filtering Accuracy

We evaluate the accuracy of a pre-alignment filter by computing its *false accept rate* and *false reject rate*. We first assess the false accept rate of Shouji, MAGNET (Alser et al., July 2017), SHD (Xin et al., 2015), and GateKeeper (Alser et al., 2017) across different edit distance thresholds and datasets. The false accept rate is the ratio of the number of dissimilar sequences that are falsely-accepted by the filter and the number of dissimilar sequences that are rejected by the optimal sequence alignment algorithm. We aim to minimize the false accept rate to maximize that number of dissimilar sequences that are eliminated. In Fig. 3, we provide the false accept rate of the four filters across our 12 datasets and edit distance thresholds of 0% to 10% of the sequence length (we provide the exact values in Section 10 in Supplementary Materials).

Based on Fig. 3, we make four key observations. (1) We observe that Shouji, MAGNET, SHD, and GateKeeper are less accurate in examining the low-edit sequences (i.e., datasets 1, 2, 5, 6, 9, and 10) than the high-edit sequences (i.e., datasets 3, 4, 7, 8, 11, and 12).

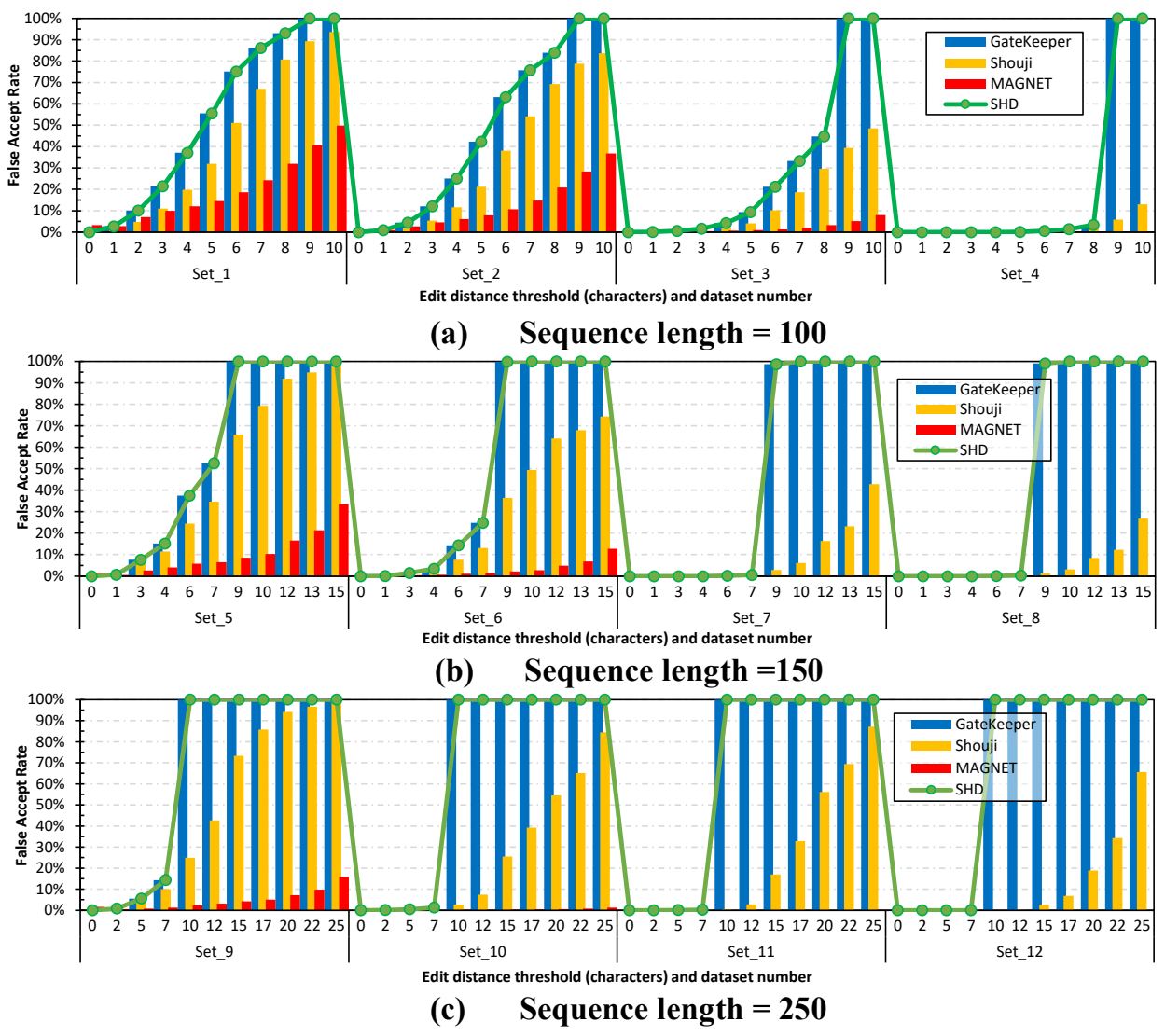

**Fig. 3: The false accept rate of Shouji, MAGNET, SHD and GateKeeper across 12 real datasets. We use a wide range of edit distance thresholds (0%-10% of the sequence length) for sequence lengths of (a) 100, (b) 150, and (c) 250.**

(2) SHD (Xin et al., 2015) and GateKeeper (Alser et al., 2017) become ineffective for edit distance thresholds of greater than 8% (*E*=8), 5% (*E*=7), and 3% (*E*=7) for sequence lengths of 100, 150, and 250 characters, respectively. This causes them to examine each sequence pair unnecessarily twice (i.e., once by GateKeeper or SHD and once by the alignment algorithm). (3) For high-edit datasets, Shouji provides up to 17.2x, 73x, and 467x (2.4x, 2.7x, and 38x for low-edit datasets) smaller false accept rate compared to GateKeeper and SHD for sequence lengths of 100, 150, and 250 characters, respectively. (4) MAGNET shows up to 1577x, 3550x, and 25552x lower false accept rates for high-edit datasets (3.5x, 14.7x, and 135x for low-edit datasets) compared to GateKeeper and SHD for sequence lengths of 100, 150, and 250 characters, respectively. MAGNET also shows up to 205x, 951x, and 16760x lower false accept rates for high-edit datasets (2.7x, 10x, and 88x for low-edit datasets) over Shouji for sequence lengths of 100, 150, and 250 characters, respectively.

We conclude that Shouji and MAGNET 1) maintain a very low rate of falsely-accepted dissimilar sequences and 2) significantly improve the accuracy of pre-alignment filtering by up to two and four orders of magnitude, respectively, compared to GateKeeper and SHD.

Second, we assess the false reject rates of pre-alignment filters in Supplementary Materials, Section 10. We demonstrate that Shouji, SHD (Xin et al., 2015) and GateKeeper (Alser et al., 2017) all have a 0% false reject rate. We also observe that MAGNET falsely-rejects correct sequence pairs, which is *unacceptable* for a reliable filter. Hence, we conclude that Shouji is the most effective pre-alignment filter, with a low false accept rate and a zero false reject rate.

## 3.3 Data Throughput and Resource Analysis

The operating frequency of our FPGA accelerator is 250 MHz. At this frequency, we observe a data throughput of nearly 3.3 GB/s, which corresponds to ~13.3 billion bases per second. This nearly reaches the peak throughput of 3.64 GB/s provided by the RIFFA (Jacobsen et al., 2015) communication channel that feeds data into the FPGA using Gen3 4-lane PCIe. We examine the FPGA resource utilization of Shouji, MAGNET, and GateKeeper (Alser et al., 2017) filters. SHD (Xin et al., 2015) is implemented in C with Intel SSE instructions and *cannot* be directly implemented on an FPGA. We examine the FPGA resource utilization for two commonly used edit distance thresholds, 2% and 5% of the sequence length, as reported in (Ahmadi et al., 2012; Alser et al., 2017; Hatem et al., 2013; Xin et al., 2015). The VC709 FPGA chip contains 433,200 slice LUTs (look-up tables) and 866,400 slice registers (flip-flops). Table 1 lists the FPGA resource utilization for a single filtering unit. We make three main observations. (1) The design for a single MAGNET filtering unit requires about 10.5% and 37.8% of the available LUTs for edit distance thresholds of 2 and 5, respectively. Hence, MAGNET can process 8 and 2 sequence pairs concurrently for edit distance thresholds of 2 and 5, respectively, without violating the timing constraints of our accelerator. (2) The design for a single Shouji filtering unit requires about 15x-21.9x fewer LUTs compared to MAGNET. This enables Shouji to achieve more parallelism over the MAGNET design as it can have 16 filtering units within the same FPGA chip. (3) GateKeeper requires about 26.9x-53x and 1.7x-2.4x fewer LUTs compared to MAGNET and Shouji, respectively. GateKeeper can also examine 16 sequence pairs at the same time.

We conclude that the FPGA resource usage is correlated with the filtering accuracy. For example, the least accurate filter, GateKeeper, occupies the least FPGA resources. Yet, Shouji has very low FPGA resource usage.

**Table 1: FPGA resource usage for a single filtering unit of Shouji, MAGNET, and GateKeeper, for a sequence length of 100 and under different edit distance thresholds. We highlight the best value in each column.**

| Filter | *E* | Single Filtering Unit | | Max. No. of |
|---|---|---|---|---|
| | | Slice LUT | Slice Register | Filtering Units |
| Shouji | 2 | 0.69% | 0.01% | 16 |
| | 5 | 1.72% | 0.01% | 16 |
| MAGNET | 2 | 10.50% | 0.8% | 8 |
| | 5 | 37.80% | 2.30% | 2 |
| GateKeeper | 2 | **0.39%** | **0.01%** | **16** |
| | 5 | 0.71% | 0.01% | 16 |

## 3.4 Filtering Speed

We analyze the execution time of MAGNET and Shouji compared to SHD (Xin et al., 2015) and GateKeeper (Alser et al., 2017). We evaluate GateKeeper, MAGNET, and Shouji using a single FPGA chip and run SHD using a single CPU core. SHD supports a sequence length of up to only 128 characters (due to the SIMD register size). To ensure as fair a comparison as possible, we allow SHD to divide the long sequences into batches of 128 characters, examine each batch individually, and then sum up the results. In Table 2, we provide the execution time of the four pre-alignment filters using 120 million sequence pairs under sequence lengths of 100 and 250 characters.

**Table 2: Execution time (in seconds) of FPGA-based GateKeeper, MAGNET, Shouji, and CPU-based SHD under different edit distance thresholds and sequence lengths. We use set_1 to set_4 for a sequence length of 100 and set_9 to set_12 for a sequence length of 250. We provide the performance results for both a single filtering unit and the maximum number of filtering units (in parentheses).**

| *E* | GateKeeper | MAGNET | Shouji | SHD |
|---|---|---|---|---|
| | | *Sequence Length = 100* | | |
| 2 | **2.89[a] (0.18[b], 16[c])** | 2.89 (0.36, 8) | **2.89 (0.18, 16)** | 60.33 |
| 5 | **2.89 (0.18, 16)** | 2.89 (1.45, 2) | **2.89 (0.18, 16)** | 67.92 |
| | | *Sequence Length = 250* | | |
| 5 | **5.78 (0.72, 8)** | 5.78 (2.89[d], 2) | **5.78 (0.72[d], 8)** | 141.09 |
| 15 | **5.78 (0.72, 8)** | 5.78 (5.78[d], 1) | **5.78 (0.72[d], 8)** | 163.82 |

[a] Execution time, in seconds, for a single filtering unit.

[b] Execution time, in seconds, for maximum filtering units.

[c] The number of filtering units.

[d] Theoretical results based on the resource utilization and data throughput.

We make four key observations. (1) Shouji's execution time is as low as that of GateKeeper (Alser et al., 2017), and 2x-8x lower than that of MAGNET. This observation is in accord with our expectation and can be explained by the fact that MAGNET has more resource overhead that limits the number of filtering units on an FPGA. Yet Shouji is up to two orders of magnitude more accurate than GateKeeper (as we show earlier in Section 3.2). (2) Shouji is up to 28x and 335x faster than SHD using one and 16 filtering units, respectively. (3) MAGNET is up to 28x and 167.5x faster than SHD using one and 8 filtering units, respectively. As we present in Supplementary Materials, Section 12, the hardware-accelerated versions of Shouji and MAGNET provide up to three orders of magnitude of speedup over their functionally-equivalent CPU implementations.

We conclude that Shouji is extremely fast and accurate. Shouji's performance also scales very well over a wide range of both edit distance thresholds and sequence lengths.

## 3.5 Effects of Pre-Alignment Filtering on Sequence Alignment

We analyze the benefits of integrating our proposed pre-alignment filter (and other filters) with state-of-the-art aligners. Table 3 presents the effect of different pre-alignment filters on the overall alignment time. We select five best-performing aligners, each of which is designed for a different type of computing platform. We use a total of 120 million real sequence pairs from our previously-described four datasets (set_1 to set_4) in this analysis. We evaluate the actual execution time of Edlib (Šošić and Šikić, 2017) and Parasail (Daily, 2016) on our machine. However, FPGASW (Fei et al., 2018), CUDASW++ 3.0 (Liu et al., 2013), and GSWABE (Liu and Schmidt, 2015) are *not* open-source and not available to us. Therefore, we scale the reported number of computed entries of the dynamic programming matrix in a second (i.e., GCUPS) as follows: 120,000,000 / (GCUPS / $100^2$). We make three key observations. (1) The execution time of Edlib (Šošić and Šikić, 2017) reduces by up to 18.8x, 16.5x, 13.9x, and 5.2x after the addition of Shouji, MAGNET, GateKeeper, and SHD, respectively, as a pre-alignment filtering step. We also observe a very similar trend for Parasail (Daily, 2016) combined with each of the four pre-alignment filters. (2) Aligners designed for FPGAs and GPUs follow a different trend than that we observe in the CPU aligners. We observe that FPGASW (Fei et al., 2018), CUDASW++ 3.0 (Liu et al., 2013), and GSWABE (Liu and Schmidt, 2015) are *faster* alone than with SHD (Xin et al., 2015) incorporated as the pre-alignment filtering step. Shouji, MAGNET, and GateKeeper (Alser et al., 2017) still significantly reduce the overall execution time of both FPGA and GPU based aligners. Shouji reduces the overall alignment time of FPGASW (Fei et al., 2018), CUDASW++ 3.0 (Liu et al., 2013), and GSWABE (Liu and Schmidt, 2015) by factors of up to 14.5x, 14.2x, and 17.9x, respectively. This is up to 1.35x, 1.4x, and 85x more than the effect of MAGNET, GateKeeper, and SHD on the end-to-end alignment time.

**Table 3: End-to-end execution time (in seconds) for several state-of-the-art sequence alignment algorithms, with and without pre-alignment filters (Shouji, MAGNET, GateKeeper, and SHD) and across different edit distance thresholds.**

| *E* | Edlib | w/ Shouji | w/ MAGNET | w/ GateKeeper | w/ SHD |
|---|---|---|---|---|---|
| 2 | 506.66 | **26.86** | 30.69 | 36.39 | 96.54 |
| 5 | 632.95 | 147.20 | **106.80** | 208.77 | 276.51 |
| ***E*** | **Parasail** | **w/ Shouji** | **w/ MAGNET** | **w/ GateKeeper** | **w/ SHD** |
| 2 | 1310.96 | **69.21** | 78.83 | 93.87 | 154.02 |
| 5 | 2044.58 | 475.08 | **341.77** | 673.99 | 741.73 |
| ***E*** | **FPGASW** | **w/ Shouji** | **w/ MAGNET** | **w/ GateKeeper** | **w/ SHD** |
| 2 | 11.33 | **0.78** | 1.04 | 0.99 | 61.14 |
| 5 | 11.33 | **2.81** | 3.34 | 3.91 | 71.65 |
| ***E*** | **CUDASW++ 3.0** | **w/ Shouji** | **w/ MAGNET** | **w/ GateKeeper** | **w/ SHD** |
| 2 | 10.08 | **0.71** | 0.96 | 0.90 | 61.05 |
| 5 | 10.08 | **2.52** | 3.13 | 3.50 | 71.24 |
| ***E*** | **GSWABE** | **w/ Shouji** | **w/ MAGNET** | **w/ GateKeeper** | **w/ SHD** |
| 2 | 61.86 | **3.44** | 4.06 | 4.60 | 64.75 |
| 5 | 61.86 | 14.55 | **11.75** | 20.57 | 88.31 |

(3) We observe that if the execution time of the aligner is much larger than that of the pre-alignment filter (which is the case for Edlib, Parasail, and GSWABE for *E*=5 characters), then MAGNET provides up to 1.3x more end-to-end speedup over Shouji. This is expected as MAGNET produces a smaller false accept rate compared to Shouji. However, unlike MAGNET, Shouji provides a 0% false reject rate. We conclude that among the four pre-alignment filters, Shouji is the best-performing pre-alignment filter in terms of both speed and accuracy. Integrating Shouji with an aligner leads to strongly positive benefits and reduces the aligner's total execution time by up to 18.8x.

## 3.6 Effects of Pre-Alignment Filtering on the Read Mapper

After confirming the benefits of integrating Shouji with sequence alignment algorithms, we now evaluate the overall benefits of integrating Shouji with the mrFAST (v. 2.6.1) mapper (Alkan et al., 2009) and BWA-MEM (Li, 2013). Table 4 summarizes the effect of Shouji on the overall mapping time, when all reads from ERR240727_1 (100 bp) are mapped to GRCh37 with an edit distance threshold of 2% and 5%. We also provide the total execution time breakdown in Table 15 in the Supplementary Materials. We make two observations. (1) The mapping time of mrFAST reduces by a factor of up to 5 after adding Shouji as the pre-alignment step. (2) Integrating Shouji with BWA-MEM, without optimizing the mapper, shows less benefit than integrating Shouji with mrFAST (up to 1.07x reduction in the overall mapping time). This is due to the fact that BWA-MEM generates a low number of pairs that require verification using the read aligner. We believe by changing the mapper to work better with Shouji, we can achieve larger speedups. We leave this for future work.

**Table 4: Overall mrFAST and BWA-MEM mapping time (in seconds) with and without Shouji, for an edit distance threshold of 2% and 5%.**

| | *E* | # pairs to be verified | # pairs rejected by Shouji | map. time w/o Shouji | mapping time w/ Shouji |
|---|---|---|---|---|---|
| mrFAST | 2 | 40,859,970 | 30,679,795 | 242.1s | 195.4s (1.2x) |
| mrFAST | 5 | 874,403,170 | 764,688,027 | 2532s | 504.6s (5.0x) |
| BWA-MEM | 2 | 653,543 | 585,036 | 668.1s | 626.9s (1.07x) |
| BWA-MEM | 2* | 8,209,193 | 7,847,125 | 670.1s | 625.8s (1.07x) |
| BWA-MEM | 5 | 660,901 | 593,247 | 695.1s | 655.8s (1.06x) |
| BWA-MEM | 5* | 8,542,937 | 8,186,550 | 696.1s | 652.7s (1.07x) |

* We configure BWA-MEM to report all secondary alignments using *-a*.

# 4 DISCUSSION AND FUTURE WORK

We demonstrate that the concept of pre-alignment filtering provides substantial benefits to the existing and future sequence alignment algorithms. Accelerated sequence aligners that offer different strengths and features are frequently introduced. Many of these efforts either simplify the scoring function, or only take into account accelerating the computation of the dynamic programming matrix *without* supporting the backtracking step. Shouji offers the ability to make the best use of existing aligners *without sacrificing any of their capabilities*, as it does *not* modify or replace the alignment step. As such, we hope that it catalyzes the adoption of specialized pre-alignment accelerators in genome sequence analysis. However, the use of specialized hardware chips may discourage users who are not necessarily fluent in FPGAs. This concern can be alleviated in at least two ways. First, the Shouji accelerator can be integrated more closely *inside* the sequencing machines to perform real-time pre-alignment filtering concurrently with sequencing (Lindner et al., 2016). This allows a significant reduction in total genome analysis time. Second, cloud computing offers access to a large number of advanced FPGA chips that can be used concurrently via a simple user-friendly interface. However, such a scenario requires the development of privacy-preserving pre-alignment filters due to privacy and legal concerns (Salinas and Li, 2017). Our next efforts will focus on exploring privacy-preserving real-time pre-alignment filtering.

Another potential target of our research is to explore the possibility of accelerating optimal alignment calculations for longer sequences (few tens of thousands of characters) (Senol et al., 2018) using pre-alignment filtering. Longer sequences pose two challenges. First, we need to transfer more data to the FPGA chip to be able process a single pair of sequences which is mainly limited by the data transfer rate of the communication link (i.e., PCIe). Second, typical edit distance threshold used for sequence alignment is 5% of the sequence length. For considerably long sequences, edit distance threshold is around few hundreds of characters. For a large edit distance threshold, each character of a given sequence is compared to a large number of neighboring characters of the other given sequence. This makes the short matches (e.g., a single zero or two consecutive zeros) to occur more frequently in the diagonal vectors, which would negatively affect the accuracy of Shouji. We will investigate this effect and explore new pre-alignment filtering approaches for the sequencing data produced by third-generation sequence machines.

## 5 CONCLUSION

In this work, we propose Shouji, a highly-parallel and accurate pre-alignment filtering algorithm accelerated on a specialized hardware platform. The key idea of Shouji is to rapidly and accurately eliminate dissimilar sequences *without* calculating banded optimal alignment. Our hardware-accelerated version of Shouji provides, on average, three orders of magnitude speedup over its functionally-equivalent CPU implementation. Shouji improves the accuracy of pre-alignment filtering by up to two orders of magnitude compared to the best-performing existing pre-alignment filter, GateKeeper. The addition of Shouji as a pre-alignment step significantly reduces the alignment time of state-of-the-art aligners by up to 18.8x, leading to the fastest alignment mechanism that we know of.

## Acknowledgments

We thank Tuan Duy Anh Nguyen for his valuable comments on the hardware design.

## Funding

This work is supported in part by the NIH Grant (HG006004 to O. Mutlu and C. Alkan) and the EMBO Installation Grant (IG-2521) to C. Alkan. M. Alser is supported in part by the HiPEAC collaboration grant and TUBITAK-2215 graduate fellowship from the Scientific and Technological Research Council of Turkey.

*Conflict of Interest:* none declared.

## References

Ahmadi, A., Behm, A., Honnalli, N., Li, C., Weng, L. and Xie, X. (2012) Hobbes: optimized gram-based methods for efficient read alignment, *Nucleic acids research*, **40**, e41-e41.

Al Kawam, A., Khatri, S. and Datta, A. (2017) A Survey of Software and Hardware Approaches to Performing Read Alignment in Next Generation Sequencing, *IEEE/ACM Transactions on Computational Biology and Bioinformatics (TCBB)*, **14**, 1202-1213.

Alkan, C., Kidd, J. M., Marques-Bonet, T., Aksay, G., Antonacci, F., Hormozdiari, F., Kitzman, J. O., Baker, C., Malig, M. and Mutlu, O. (2009) Personalized copy number and segmental duplication maps using next-generation sequencing, *Nature genetics*, **41**, 1061-1067.

Alser, M., Hassan, H., Xin, H., Ergin, O., Mutlu, O. and Alkan, C. (2017) GateKeeper: a new hardware architecture for accelerating pre-alignment in DNA short read mapping, *Bioinformatics*, **33**, 3355-3363.

Alser, M., Mutlu, O. and Alkan, C. (July 2017) Magnet: Understanding and improving the accuracy of genome pre-alignment filtering, *Transactions on Internet Research* **13**.

Aluru, S. and Jammula, N. (2014) A review of hardware acceleration for computational genomics, *Design & Test, IEEE*, **31**, 19-30.

Backurs, A. and Indyk, P. (2017) Edit Distance Cannot Be Computed in Strongly Subquadratic Time (unless SETH is false), *arXiv preprint arXiv:1412.0348v4*

Banerjee, S. S., El-Hadedy, M., Lim, J. B., Kalbarczyk, Z. T., Chen, D., Lumetta, S. and Iyer, R. K. (2018) ASAP: Accelerated Short-Read Alignment on Programmable Hardware, *arXiv preprint arXiv:1803.02657*.

Calude, C., Salomaa, K. and Yu, S. (2002) Additive distances and quasi-distances between words, *Journal of Universal Computer Science*, **8**, 141-152.

Chen, P., Wang, C., Li, X. and Zhou, X. (2014) Accelerating the next generation long read mapping with the FPGA-based system, *IEEE/ACM Transactions on Computational Biology and Bioinformatics (TCBB)*, **11**, 840-852.

Chen, Y.-T., Cong, J., Fang, Z., Lei, J. and Wei, P. (2016) When spark meets FPGAs: a case study for next-generation DNA sequencing acceleration. *Field-Programmable Custom Computing Machines (FCCM), 2016 IEEE 24th Annual International Symposium on*. IEEE, pp. 29-29.

Consortium, G. P. (2012) An integrated map of genetic variation from 1,092 human genomes, *Nature*, **491**, 56-65.

Daily, J. (2016) Parasail: SIMD C library for global, semi-global, and local pairwise sequence alignments, *BMC bioinformatics*, **17**, 81.

Fei, X., Dan, Z., Lina, L., Xin, M. and Chunlei, Z. (2018) FPGASW: Accelerating Large-Scale Smith–Waterman Sequence Alignment Application with Backtracking on FPGA Linear Systolic Array, *Interdisciplinary Sciences: Computational Life Sciences*, **10**, 176-188.

Fox, E. J., Reid-Bayliss, K. S., Emond, M. J. and Loeb, L. A. (2014) Accuracy of next generation sequencing platforms, *Next generation, sequencing & applications*, **1**.

Georganas, E., Buluç, A., Chapman, J., Oliker, L., Rokhsar, D. and Yelick, K. (2015) meraligner: A fully parallel sequence aligner. *Parallel and Distributed Processing Symposium (IPDPS), 2015 IEEE International*. IEEE, pp. 561-570.

Hatem, A., Bozdağ, D., Toland, A. E. and Çatalyürek, Ü. V. (2013) Benchmarking short sequence mapping tools, *BMC bioinformatics*, **14**, 184.

Henikoff, S. and Henikoff, J. G. (1992) Amino acid substitution matrices from protein blocks, *Proceedings of the National Academy of Sciences*, **89**, 10915-10919.

Herbordt, M. C., VanCourt, T., Gu, Y., Sukhwani, B., Conti, A., Model, J. and DiSabello, D. (2007) Achieving high performance with FPGA-based computing, *Computer*, **40**, 50.

Jacobsen, M., Richmond, D., Hogains, M. and Kastner, R. (2015) RIFFA 2.1: A Reusable Integration Framework for FPGA Accelerators, *ACM Trans. Reconfigurable Technol. Syst.*, **8**, 1-23.

Kim, J. S., Cali, D. S., Xin, H., Lee, D., Ghose, S., Alser, M., Hassan, H., Ergin, O., Alkan, C. and Mutlu, O. (2018) GRIM-Filter: Fast seed location filtering in DNA read mapping using processing-in-memory technologies, *BMC genomics*, **19**, 89.

Kung, H.-T. (1982) Why systolic architectures?, *IEEE computer*, **15**, 37-46.

Levenshtein, V. I. (1966) Binary codes capable of correcting deletions, insertions, and reversals, *Soviet physics doklady*, **10**.

Li, H. (2013) Aligning sequence reads, clone sequences and assembly contigs with BWA-MEM, *arXiv preprint arXiv:1303.3997*.

Lindner, M. S., Strauch, B., Schulze, J. M., Tausch, S., Dabrowski, P. W., Nitsche, A. and Renard, B. Y. (2016) HiLive–Real-Time Mapping of Illumina Reads while Sequencing, *Bioinformatics*, btw659.

Lipman, D. J. and Pearson, W. R. (1985) Rapid and sensitive protein similarity searches, *Science*, **227**, 1435-1441.

Liu, Y. and Schmidt, B. (2015) GSWABE: faster GPU-accelerated sequence alignment with optimal alignment retrieval for short DNA sequences, *Concurrency and Computation: Practice and Experience*, **27**, 958-972.

Liu, Y., Wirawan, A. and Schmidt, B. (2013) CUDASW++ 3.0: accelerating Smith-Waterman protein database search by coupling CPU and GPU SIMD instructions, *BMC bioinformatics*, **14**, 117.

Masek, W. J. and Paterson, M. S. (1980) A faster algorithm computing string edit distances, *Journal of Computer and System Sciences*, **20**, 18-31.

McKernan, K. J., Peckham, H. E., Costa, G. L., McLaughlin, S. F., Fu, Y., Tsung, E. F., Clouser, C. R., Duncan, C., Ichikawa, J. K. and Lee, C. C. (2009) Sequence and structural variation in a human genome uncovered by short-read, massively parallel ligation sequencing using two-base encoding, *Genome research*, **19**, 1527-1541.

Navarro, G. (2001) A guided tour to approximate string matching, *ACM computing surveys (CSUR)*, **33**, 31-88.

Needleman, S. B. and Wunsch, C. D. (1970) A general method applicable to the search for similarities in the amino acid sequence of two proteins, *Journal of molecular biology*, **48**, 443-453.

Ng, H.-C., Liu, S. and Luk, W. (2017) Reconfigurable acceleration of genetic sequence alignment: A survey of two decades of efforts. *Field Programmable Logic and Applications (FPL), 2017 27th International Conference on*. IEEE, pp. 1-8.

Nishimura, T., Bordim, J. L., Ito, Y. and Nakano, K. (2017) Accelerating the Smith-Waterman Algorithm Using Bitwise Parallel Bulk Computation Technique on GPU. *Parallel and Distributed Processing Symposium Workshops (IPDPSW), 2017 IEEE International*. IEEE, pp. 932-941.

Salinas, S. and Li, P. (2017) Secure Cloud Computing for Pairwise Sequence Alignment. *Proceedings of the 8th ACM International Conference on Bioinformatics, Computational Biology, and Health Informatics*. ACM, pp. 178-183.

Sandes, E. F. D. O., Boukerche, A. and Melo, A. C. M. A. D. (2016) Parallel optimal pairwise biological sequence comparison: Algorithms, platforms, and classification, *ACM Computing Surveys (CSUR)*, **48**, 63.

Senol, C. D., Kim, J., Ghose, S., Alkan, C. and Mutlu, O. (2018) Nanopore sequencing technology and tools for genome assembly: computational analysis of the current state, bottlenecks and future directions, *Briefings in bioinformatics*.

Seshadri, V., Lee, D., Mullins, T., Hassan, H., Boroumand, A., Kim, J., Kozuch, M. A., Mutlu, O., Gibbons, P. B. and Mowry, T. C. (2017) Ambit: In-memory accelerator for bulk bitwise operations using commodity DRAM technology. *Proceedings of the 50th Annual IEEE/ACM International Symposium on Microarchitecture*. ACM, pp. 273-287.

Smith, T. F. and Waterman, M. S. (1981) Identification of common molecular subsequences, *Journal of molecular biology*, **147**, 195-197.

Šošić, M. and Šikić, M. (2017) Edlib: a C/C++ library for fast, exact sequence alignment using edit distance, *Bioinformatics*, **33**, 1394-1395.

Trimberger, S. M. (2015) Three ages of FPGAs: a retrospective on the first thirty years of FPGA technology, *Proceedings of the IEEE*, **103**, 318-331.

Ukkonen, E. (1985) Algorithms for approximate string matching, *Information and control*, **64**, 100-118.

Waidyasooriya, H. and Hariyama, M. (2015) Hardware-Acceleration of Short-read Alignment Based on the Burrows-Wheeler Transform, *Parallel and Distributed Systems, IEEE Transactions on*, **PP**, 1-1.

Wang, C., Yan, R.-X., Wang, X.-F., Si, J.-N. and Zhang, Z. (2011) Comparison of linear gap penalties and profile-based variable gap penalties in profile–profile alignments, *Computational biology and chemistry*, **35**, 308-318.

Xilinx (2014) Virtex-7 XT VC709 Connectivity Kit, Getting Started Guide, **UG966 (v3.0.1) June 30, 2014**.

Xin, H., Greth, J., Emmons, J., Pekhimenko, G., Kingsford, C., Alkan, C. and Mutlu, O. (2015) Shifted Hamming Distance: A Fast and Accurate SIMD-Friendly Filter to Accelerate Alignment Verification in Read Mapping, *Bioinformatics*, **31**, 1553-1560.

Xin, H., Lee, D., Hormozdiari, F., Yedkar, S., Mutlu, O. and Alkan, C. (2013) Accelerating read mapping with FastHASH, *BMC genomics*, **14**, S13.

## Supplementary Materials

# 6 Shouji Filter

### 6.1 Examining the Effect of Different Window Sizes on the Accuracy of the Shouji Algorithm.

In Fig. 4, we experimentally evaluate the effect of different window sizes on the false accept rate of Shouji. We observe that as we increase the window size, the rate of dissimilar sequences that are accepted by Shouji decreases. This is because individual matches (i.e., single zeros) are usually useless and they are not necessarily part of the common subsequences. As we increase the search window size, we are ignoring these individual matches and instead we only look for longer streaks of consecutive zeros. We also observe that a window size of 4 columns provides the lowest false accept rate (i.e., the highest accuracy).

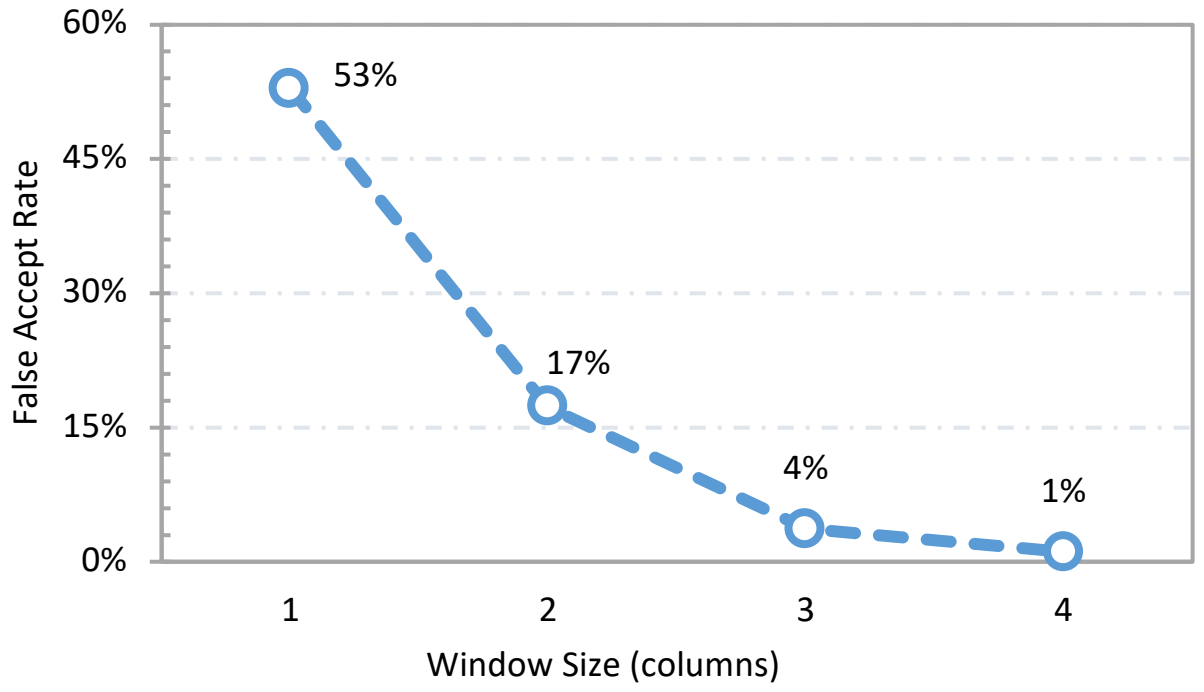

**Fig. 4: The effect of the window size on the rate of the falsely-accepted sequences (i.e., dissimilar sequences that are considered as similar ones by Shouji filter). We observe that a window width of 4 columns provides the highest accuracy. We also observe that as window size increases beyond 4 columns, more similar sequences are rejected by Shouji, which should be avoided.**

### 6.2 The Shouji Algorithm and Its Analysis

We provide the Shouji algorithm along with analysis of its computational complexity (asymptotic run time and space complexity). Shouji divides the problem of finding the common subsequences into at most *m* subproblems, as described in Algorithm 1 (line 9). Each subproblem examines each of the 2*E*+1 bit-vectors and finds the 4-bit subsequence that has the largest number of zeros within the sliding window (line 13 to line 23). Once found, Shouji also compares the found subsequence with its corresponding subsequence in the Shouji bit-vector and stores the subsequence that has more zeros in the Shouji bit-vector (line 24). Now, let *c* be a constant representing the run time of examining a subsequence of 4 bits long. Then, the time complexity of the Shouji algorithm is as follows:

$$T_{Shouji}(m) = c.m.(2E+2) \quad (2)$$

This demonstrates that the Shouji algorithm runs in linear time with respect to the sequence length and edit distance threshold. The Shouji algorithm maintains 2*E*+1 diagonal bit-vectors and an additional auxiliary bit-vector (i.e., the Shouji bit-vector) for each two given sequences. The space complexity of the Shouji algorithm is as follows:

$$D_{Shouji}(m) = m.(2E+2) \quad (3)$$

Hence, the Shouji algorithm requires linear space with respect to the sequence length and edit distance threshold. Next, we describe the hardware implementation details of the Shouji filter.

### 6.3 Hardware Implementation

We present the FPGA chip layout for our hardware accelerator in Fig. 5. As we illustrated in the main manuscript, Section 2.3, we implement the first step of our Shouji algorithm, building neighborhood map, using shift registers and bitwise XOR operations. The second step of the Shouji algorithm is identifying the diagonally-consecutive matches. This key step involves finding the 4-bit vector that has the largest number of zeros. For each search window, there are 2*E*+1 diagonal bit-vectors and an additional Shouji bit-vector. To enable the computation to be performed in a parallel fashion, we build 2*E*+2 counters. As presented in Fig. 5, each counter counts the number of zeros in a single bit-vector. The counter takes four bits as input and generates three bits that represent the number of zeros within the window. Each counter requires three 4-input LUTs, as each LUT has a single output signal. In total, we need *6E+6* 4-input LUTs to build a single search window. All bits of the counter output are generated at the same time, as the propagation delay through an FPGA look-up table is independent of the implemented function (Xilinx, November 17, 2014). The comparator is responsible for selecting the 4-bit subsequence that maximizes the number of consecutive matches based on the output of each counter and the Shouji bit-vector. Finally, the selected 4-bit subsequence is then stored in the Shouji bit-vector at the same corresponding location.

**Algorithm 1**: Shouji — Comments

**Input**: text (*T*), pattern (*P*), *edit distance threshold* (*E*).
**Output**: *1* (*Similar/Alignment is needed*) / *0* (*Dissimilar/Alignment is not needed*).

```
m ← length(T);
for i ← 1 to m do
    for j ← i-E to i+E do
        if T[i] == P[j] then
            N[i,j] ← 0;
        else N[i,j]← 1;
```

**Step 1**: Building neighborhood map (*N*)

Output: 2*E*+1 diagonal bit-vectors

```
for i ← 1 to m do Shouji[i] ← 1; //initializing Shouji bit-vector to 1's
Z ← [0000]; // Z is 4-bit vector that stores the longest streak of diagonally-consecutive zeros
for i ← 1 to m do // slide the search window by a single step
    for j ← 1 to E do // iterate over the diagonals
        // function CZ(D) counts the occurrence of zeros in its input bit-vector D
        // Compare jth lower diagonal with jth upper diagonal
        if CZ(N[i+j:i+3+j,i:i+3]) > CZ(N[i:i+3,i+j:i+3+j]) then
            Z ← N[i+j:i+3+j,i:i+3];
        // If jth lower and jth upper diagonals have the same number of
        // zeros then selects the diagonal that starts with zeros
        else if CZ(N[i+j:i+3+j,i:i+3]) == CZ(N[i:i+3,i+j:i+3+j]) then
            if N[i+j,i]==0 then Z ← N[i+j:i+3+j,i:i+3];
            else if N[i,i+j]==0 then Z ← N[i:i+3,i+j:i+3+j];
        // Compare Z with the jth upper diagonal
        else Z ← N[i:i+3,i+j:i+3+j];
    // Compare Z with main diagonal and Shouji bit-vector
    if CZ(N[i:i+3,i:i+3]) > CZ(Z) then  Z ← N[i:i+3,i:i+3];
    if CZ(Z) > CZ(Shouji[i:i+3]) then  Shouji[i:i+3] ← Z;
```

**Step 2**: Identifying the Diagonally-Consecutive Matches

```
if CZ(Shouji) ≥ m-E then return 1;
else return 0;
```

**Step 3**: Filtering out Dissimilar Sequences

**Algorithm 2**: CZ (count zeros) function

**Function**: CZ() counts the number of occurrences of zeros.
**Input**: bit-vector *D*.
**Output**: number of occurrences of zeros.

```
count ← 0;
for i ← 1 to length(D) do
    if D[i] == 0 then
        count ← count + 1;
return count;
```

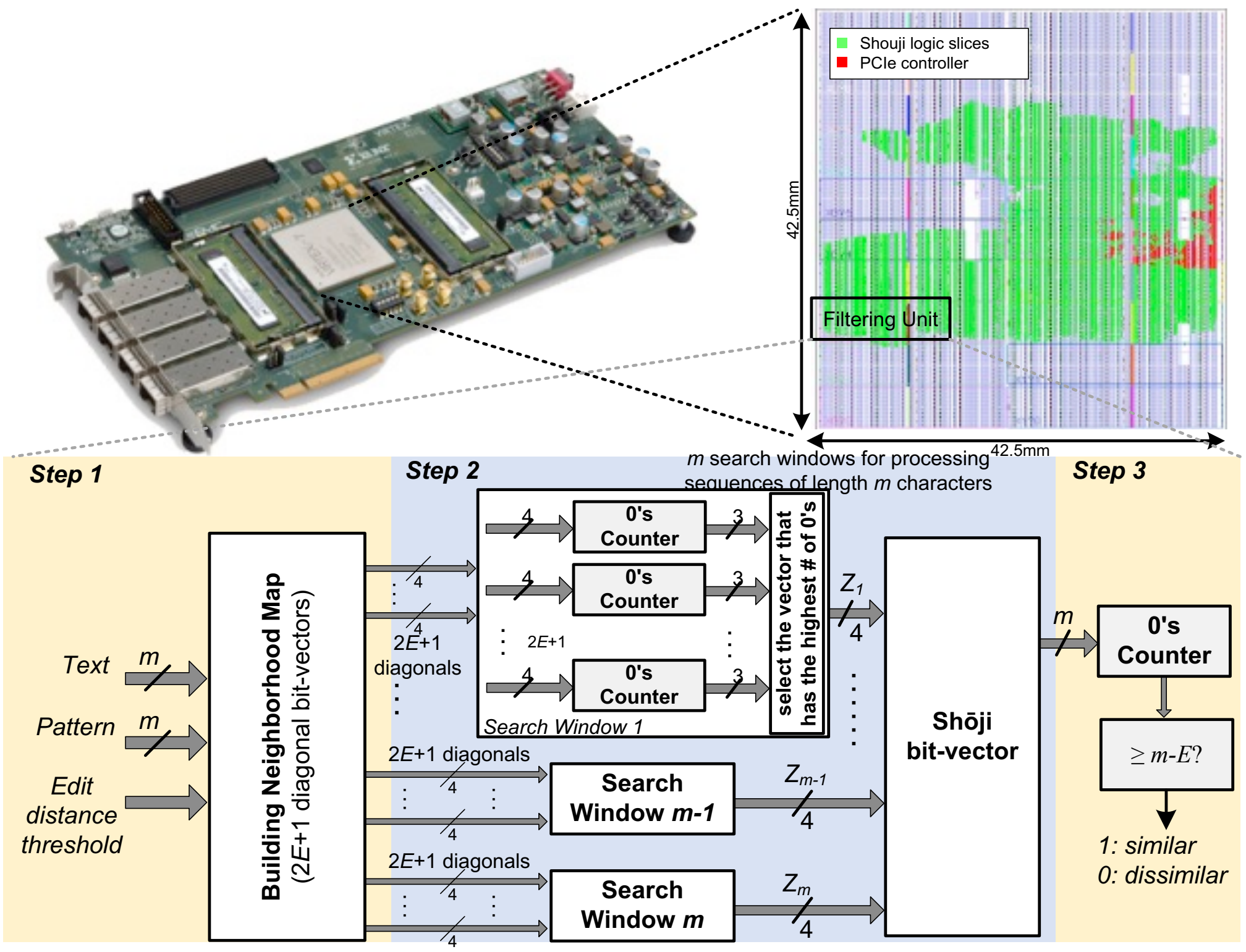

**Fig. 5: FPGA chip layout for Shouji and block diagram of the search window scheme implemented in a Xilinx VC709 FPGA for a single filtering unit.**

## 7 MAGNET Filter

First, we provide the MAGNET (Alser et al., July 2017) algorithm and describe its main filtering mechanism. Second, we analyze the computational complexity of the MAGNET algorithm. Third, we provide details about the hardware implementation of the MAGNET algorithm.

### 7.1 Overview

MAGNET (Alser et al., July 2017) is another filter that uses a divide-and-conquer technique to find all the $E+1$ common subsequences, if any, and sum up their length. By calculating their total length, we can estimate the total number of edits between the two given sequences. If the total length of the $E+1$ common subsequences is less than $m$-$E$, then there exist more common subsequences than $E+1$ that are associated with more edits than allowed. If so, then MAGNET excludes the two given sequences from optimal alignment calculation. We present the algorithm of MAGNET in Algorithm 3.

| **Algorithm 3**: MAGNET | Comments |
|---|---|
| **Input**: text (*T*), pattern (*P*), *edit distance threshold* (*E*).<br>**Output**: *1* (*Similar/Alignment is needed*) / *0* (*Dissimilar/Alignment is not needed*). | |
| 1: *m* ← *length(T)*;<br>2: **for** *i* ← *1 to m* **do**<br>3: **for** *j* ← *i-E to i+E* **do**<br>4: **if** *T*[*i*] == *P*[*j*] **then**<br>5: *N*[*i,j*] ← *0*;<br>6: **else** *N*[*i,j*]← *1*; | **Step 1**: Building neighborhood map (*N*)<br><br>Output: 2*E*+1 diagonal bit-vectors |
| 7: **for** *i* ← *1 to m* **do**<br>8: *MAGNET[i]* ← *1*; *// Initializing MAGNET bit-vector*<br>9: *[MAGNET, calls]* ← *EXEN(N, 1, m, E, MAGNET, 1)*; | **Step 2 - Step 4** |
| 10: *// Function CZ() returns number of zeros*<br>11: **if** *CZ(MAGNET)* ≥ *m-E* **then** *return 1*; **else** *return 0*; | **Step 5**: Filtering out Dissimilar Sequences |

Finding the common subsequences involves four main steps. (1) **Building the neighborhood map**. Similar to Shouji, MAGNET starts with building the 2$E$+1 diagonal bit-vectors of the neighborhood map for the two given sequences (Algorithm 3, lines 2-6). (2) **Extraction.** Each diagonal bit-vector nominates its local longest subsequence of consecutive zeros. Among all nominated subsequences, a single subsequence is selected as a global longest subsequence based on its length (Algorithm 4, lines 2-11). MAGNET evaluates if the length of the global longest subsequence is less than $\lceil(m-E)/(E+1)\rceil$, then the two sequences contain more edits than allowed, which cause the common subsequences to be shorter (i.e., each edit results in dividing the sequence pair into more common subsequences). If so, then the two sequences are rejected (Algorithm 4, lines 12-13). Otherwise, MAGNET stores the length of the global longest subsequence to be used towards calculating the total length of all $E+1$ common subsequences. The lower bound equality occurs when all edits are equispaced and all $E+1$ subsequences are of the same length. (3) **Encapsulation.** The next step is essential to preserve the original edit (or edits) that causes a single common sequence to be divided into smaller subsequences. MAGNET penalizes the found subsequence by two edits (one for each side). This is achieved by excluding from the search space of all bit-vectors the indices of the found subsequence in addition to the index of the surrounding single bit from both left and right sides (Algorithm 4, lines 14-17). (4) **Divide-and-Conquer Recursion.** In order to locate the other $E$ non-overlapping subsequences, MAGNET applies a divide-and-conquer technique where we decompose the problem of finding the non-overlapping common subsequences into two subproblems. While the first subproblem focuses on finding the next long subsequence that is located on the right-hand side of the previously found subsequence in the first *extraction* step (Algorithm 4, line 15), the second subproblem focuses on the other side of the found subsequence (Algorithm 4, line 17). Each subproblem is solved by recursively repeating all the three steps mentioned above, but without evaluating again the length of the longest subsequence. MAGNET applies two early termination methods that aim to reduce the execution time of the filter. The first method is evaluating the length of the longest subsequence in the first recursion call (Algorithm 4, lines 12-13). The second method is limiting the number of the subsequences to be found to at most $E+1$, regardless of their actual number for the given sequence pair (Algorithm 4, line 1). (5) **Filtering out Dissimilar Sequences.** Once after the termination, if the total length of all found common subsequences is less than $m$-$E$, then the two sequences are rejected. Otherwise, they are considered to be similar and the alignment can be measured using sophisticated alignment algorithms.

| **Algorithm 4**: EXEN function | Comments |
|---|---|
| **Function**: EXEN() extracts the longest subsequence of consecutive zeros and generate two subproblems. | |
| **Input**: Neighborhood map (*N*), start index (*SI*), end index (*EI*), *E*, MAGNET bit-vector, number of recursion calls. | |
| **Output**: updated MAGNET bit-vector, updated number of calls. | |
| 1: **if** (SI ≤ EI and calls ≤ E+1) **then** *// Early termination condition* | |
| 2: *// Function CCZ() returns number and indices of longest* | **Step 2**: Extracting the longest subsequence of consecutive zeros |
| 3: *// subsequence of diagonally consecutive zeros* | |
| 4: **for** *j ← 1 to E* **do** *//Extraction* | |
| 5: *[X,s1,e1] ← CCZ(N[SI+j,SI],EI); // Lower diagonal* | |
| 6: *[Y,s2,e2] ← CCZ(N[SI,SI+j],EI); // Upper diagonal* | |
| 7: **if** *X > Y* **then** *s ← s1*; *e ← e1*; | |
| 8: **else** *s ← s2*; *e ← e2*; | |
| 9: *[X,s1,e1] ← CCZ(N[SI,SI],EI)*; | |
| 10: **if** *X > (e-s+1)* **then** | |
| 11: *s ← s1*; *e ← e1*; | |
| 12: **if** *(calls=1 and (e-s+1)<*$\lceil(m-E)/(E+1)\rceil$*)* **then** | Early termination condition (only in first call) |
| 13: **return** *[MAGNET, 0]*; | |
| 14: *// Right subproblem with encapsulation* | **Step 3**: Encapsulating the found longest subsequence and **Step 4**: Divide-and-Conquer Recursion |
| 15: *[MAGNET, calls] ← EXEN(N,e+2,EI, E,MAGNET, calls+1)*; | |
| 16: *// Left subproblem with encapsulation* | |
| 17: *[MAGNET, calls] ← EXEN(N,SI, s-2, E, MAGNET, calls+1)*; | |
| 18: **return** *[MAGNET, calls]*; | |
| 19: **else return** *[MAGNET, calls-1]*; | |

## 7.2 Analysis of the MAGNET Algorithm

We analyze the asymptotic run time and space complexity of the MAGNET algorithm. MAGNET applies a divide-and-conquer technique that divides the problem of finding the common subsequences into two subproblems in each recursion call. In the first recursion call, the extracted common subsequence is of length at least $a=\lceil(m-E)/(E+1)\rceil$ bases. This reduces the problem of finding the common subsequences from $m$ to at most $m$-$a$, which is further divided into two subproblems: a left subproblem and a right subproblem. For the sake of simplicity, we assume that the size of the left and the right subproblems decreases by a factor of $b$ and $c$, respectively, as follows:

$$m = a + 2 + m/b + m/c \tag{4}$$

The addition of 2 bases is for the encapsulation bits added at each recursion call. Now, let $T_{MAGNET}(m)$ be the time complexity of MAGNET algorithm, for identifying non-overlapping subsequences. If it takes $O(km)$ time to find the global longest subsequence and divide the problem into two subproblems, where $k = 2E+1$ is the number of bit-vectors, we get the following recurrence equation:

$$T_{MAGNET}(m) = T_{MAGNET}(m/b) + T_{MAGNET}(m/c) + O(km) \tag{5}$$

Given that the early termination condition of MAGNET algorithm restricts the recursion depth as follows:

$$\textit{Recursion tree depth} = \lceil log_2(E+1)\rceil - 1 \tag{6}$$

Solving the recurrence in (5) using (4) and (6) by applying the recursion-tree method provides a loose upper-bound to the time complexity as follows:

$$T_{MAGNET}(m) = O(km)\,.\sum_{x=0}^{\lceil log_2(E+1)\rceil-1}\left(\frac{1}{b}+\frac{1}{c}\right)^x$$

$$\approx O(fkm) \tag{7}$$

where $f$ is a fractional number satisfies the following range: $1 \leq f < 2$. This in turn demonstrates that the MAGNET algorithm runs in linear time with respect to the sequence length and edit distance threshold and hence it is computationally inexpensive. The space complexity of the MAGNET algorithm is as follows:

$$D_{MAGNET}(m) = D_{MAGNET}(m/b) + D_{MAGNET}(m/c) + (km+m) \approx O(fkm + fm) \quad (8)$$

Hence, MAGNET algorithm requires linear space with respect to the read length and edit distance threshold. Next, we describe the hardware implementation details of MAGNET filter.

### 7.3 Hardware Implementation

We outline the challenges that are encountered in implementing the MAGNET filter to be used in our accelerator design. Implementing the MAGNET algorithm on an FPGA is more challenging than implementing the Shouji algorithm due to the random location and variable length of each of the $E$+1 common subsequences. Verilog-2011 imposes two challenges on our architecture as it does not support variable-size partial selection and indexing of a group of bits from a vector (McNamara, 2001). In particular, the first challenge lies in excluding the extracted common subsequence along with its encapsulation bits from the search space of the next recursion call. The second challenge lies in dividing the problem into two subproblems, each of which has an unknown size at design time. To address these limitations and tackle the two design challenges, we keep the problem size fixed at each recursion call. We exclude the longest found subsequence from the search space by amending all bits of all $2E+1$ bit-vectors that are located within the indices (locations) of the encapsulation bits to '1's. This ensures that we exclude the longest found subsequence and its corresponding location in all other bit-vectors during the subsequent recursion calls. We build the MAGNET accelerator using the same FPGA board as that used for Shouji for a fair comparison.

## 8 Examples of Applying the Shouji and MAGNET algorithms

In this section, we provide three examples of applying the Shouji and MAGNET filtering algorithms to different sequence pairs. In Fig. 6, we set the edit distance threshold to 4 in these examples. The diagonal vectors of the neighborhood map are horizontally presented in the same order of the diagonal vectors for a better illustration. In the first two examples (Fig. 6(a) and Fig. 6(b)), we observe that MAGNET is highly accurate in providing the exact location of the edits in the MAGNET bit-vector. This is due to two main reasons. First, MAGNET finds the exact length of each common subsequence by performing multiple individual iteration for each common subsequence. Second, it manually encapsulates each found longest subsequence of consecutive zeros by ones, which ensures to maintain the edits in the MAGNET bit-vector. On the contrary, Shouji uses overlapping search windows to detect segments of consecutive zeros. If two segments of consecutive zeros are overlapped within a single search window, then the edit between the two segments is sometimes eliminated by the overlapping zeros of the two segments as shown in Fig. 6(a).

Pairwise alignment can be performed as a *global* alignment, where two sequences of the same length are aligned end-to-end, or a *local* alignment, where subsequences of the two given sequences are aligned. It can also be performed as a *semi-global* alignment (called glocal), where the entirety of one sequence is aligned towards one of the ends of the other sequence. To ensure correct pre-alignment filtering and avoid rejecting a correct alignment, pre-alignment filter needs to consider counting the number of edits in a similar way to that of optimal alignment algorithm. This means that if the optimal alignment algorithm performs local alignment, then the pre-alignment filter should also perform local edit distance calculation. This can be achieved by not considering the leading and trailing edits in the total count of edits between two given sequences. Fig 6(a) and Fig. 6(b) show examples of global pre-alignment filtering. Fig 6(c) shows an example of local pre-alignment filtering, where the two given sequences have different lengths. While Shouji is conceptually able to perform local pre-alignment and glocal pre-alignment filtering, such support is not currently implemented in our public release of Shouji (https://github.com/CMU-SAFARI/Shouji). The current implementation of Shouji performs only global pre-alignment filtering that requires the text and reference sequences to be of the same length.

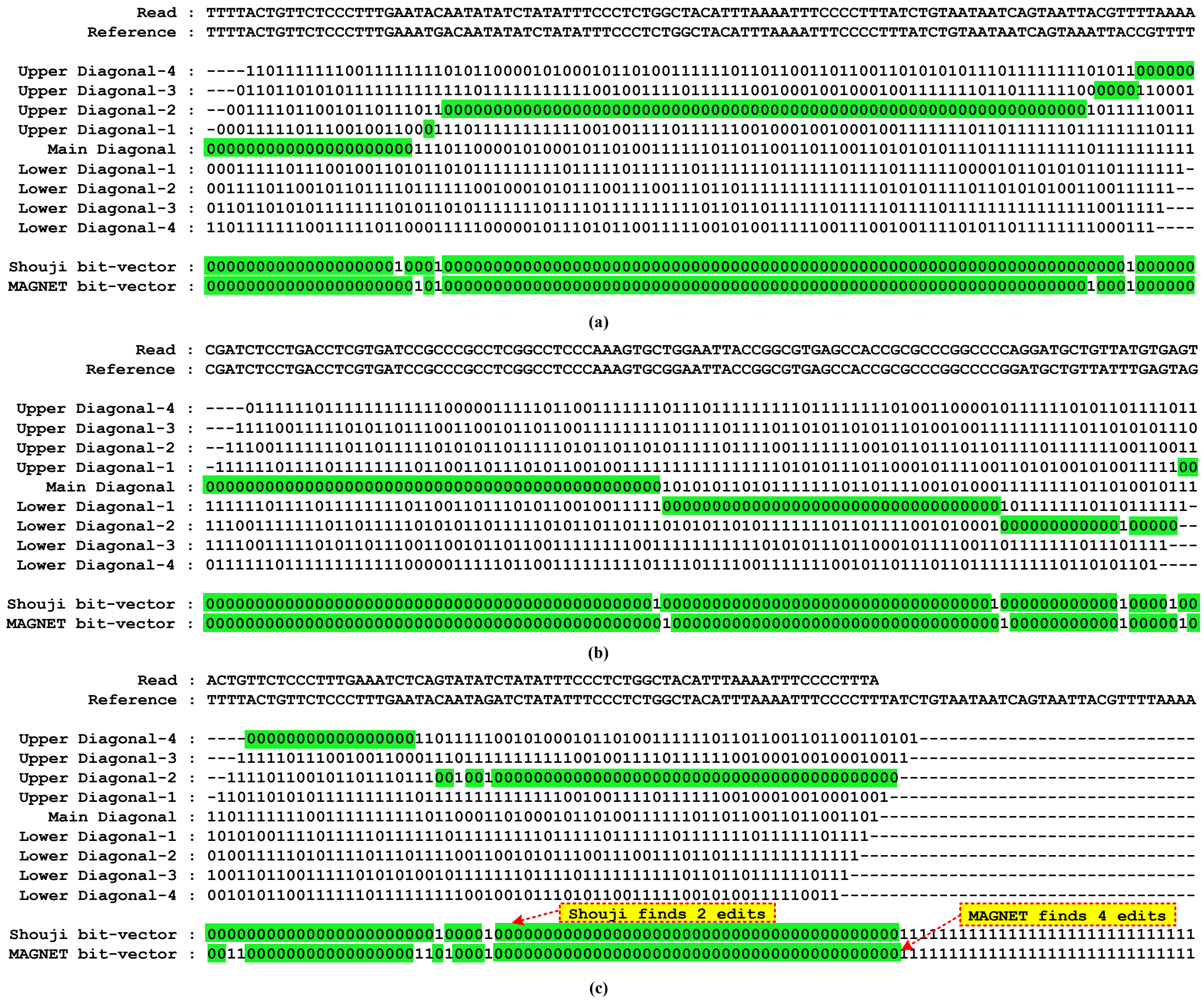

**Fig. 6: Examples of applying the Shouji and MAGNET filtering algorithms to three different sequence pairs, where the edit distance threshold is set to 4. We present the content of the neighborhood map along with the Shouji and MAGNET bit-vectors. In (a) and (b), we apply Shouji and MAGNET algorithms starting from the leftmost column towards the rightmost column (end-to-end) to perform global pre-alignment filtering. In (c), we ignore the ones that are located at the two ends of the final bit-vector to perform local pre-alignment filtering.**

## 9 Dataset Description

Table 5 provides the configuration used for the *-e* parameter of mrFAST (Alkan et al., 2009) for each of the 12 datasets. We use Edlib (Šošić and Šikić, 2017) to assess the number of similar (i.e., having edits fewer than or equal to the edit distance threshold) and dissimilar (i.e., having more edits than the edit distance threshold) pairs for each of the 12 datasets across different user-defined edit distance thresholds. We provide these details for set 1, set 2, set 3, and set 4 in Table 6. We provide the same details for set 5, set 6, set 7, and set 8 in Table 7 and for set 9, set 10, set 11, and set 12 in Table 8.

**Table 5: Benchmark illumina-like datasets (read-reference pairs). We map each read set to the human reference genome in order to generate four datasets using different mappers' edit distance thresholds (using the *-e* parameter).**

| **Accession no.** | **ERR240727_1** | | | | **SRR826460_1** | | | | **SRR826471_1** | | | |
|---|---|---|---|---|---|---|---|---|---|---|---|---|
| **Sequence Length** | 100 | | | | 150 | | | | 250 | | | |
| **HTS** | Illumina HiSeq 2000 | | | | Illumina HiSeq 2000 | | | | Illumina HiSeq 2000 | | | |
| **Dataset** | Set_1 | Set_2 | Set_3 | Set_4 | Set_5 | Set_6 | Set_7 | Set_8 | Set_9 | Set_10 | Set_11 | Set_12 |
| **mrFAST *-e*** | 2 | 3 | 5 | 40 | 4 | 6 | 10 | 70 | 8 | 12 | 15 | 100 |
| **Amount of Edits** | Low-edit | | High-edit | | Low-edit | | High-edit | | Low-edit | | High-edit | |

**Table 6: Details of our first four datasets (set 1, set 2, set 3, and set 4). We use Edlib to benchmark the accepted (i.e., aligned) pairs and the rejected (i.e., unaligned) pairs for edit distance thresholds of *E*=0 up to *E*=10 edits.**

| Dataset | Set_1 | | Set_2 | | Set_3 | | Set_4 | |
|---|---|---|---|---|---|---|---|---|
| *E* | Accepted | Rejected | Accepted | Rejected | Accepted | Rejected | Accepted | Rejected |
| 0 | 381,901 | 29,618,099 | 124,531 | 29,875,469 | 11,989 | 29,988,011 | 11 | 29,999,989 |
| 1 | 1,345,842 | 28,654,158 | 441,927 | 29,558,073 | 44,565 | 29,955,435 | 18 | 29,999,982 |
| 2 | 3,266,455 | 26,733,545 | 1,073,808 | 28,926,192 | 108,979 | 29,891,021 | 24 | 29,999,976 |
| 3 | 5,595,596 | 24,404,404 | 2,053,181 | 27,946,819 | 206,903 | 29,793,097 | 27 | 29,999,973 |
| 4 | 7,825,272 | 22,174,728 | 3,235,057 | 26,764,943 | 334,712 | 29,665,288 | 29 | 29,999,971 |
| 5 | 9,821,308 | 20,178,692 | 4,481,341 | 25,518,659 | 490,670 | 29,509,330 | 34 | 29,999,966 |
| 6 | 11,650,490 | 18,349,510 | 5,756,432 | 24,243,568 | 675,357 | 29,324,643 | 83 | 29,999,917 |
| 7 | 13,407,801 | 16,592,199 | 7,091,373 | 22,908,627 | 891,447 | 29,108,553 | 177 | 29,999,823 |
| 8 | 15,152,501 | 14,847,499 | 8,531,811 | 21,468,189 | 1,151,447 | 28,848,553 | 333 | 29,999,667 |
| 9 | 16,894,680 | 13,105,320 | 10,102,726 | 19,897,274 | 1,469,996 | 28,530,004 | 711 | 29,999,289 |
| 10 | 18,610,897 | 11,389,103 | 11,807,488 | 18,192,512 | 1,868,827 | 28,131,173 | 1,627 | 29,998,373 |

**Table 7: Details of our second four datasets (set_5, set_6, set_7, and set_8). We report the accepted and the rejected pairs for edit distance thresholds of *E*=0 up to *E*=15 edits.**

| Dataset | Set_5 | | Set_6 | | Set_7 | | Set_8 | |
|---|---|---|---|---|---|---|---|---|
| *E* | Accepted | Rejected | Accepted | Rejected | Accepted | Rejected | Accepted | Rejected |
| 0 | 1,440,497 | 28,559,503 | 248,920 | 29,751,080 | 444 | 29,999,556 | 201 | 29,999,799 |
| 1 | 1,868,909 | 28,131,091 | 324,056 | 29,675,944 | 695 | 29,999,305 | 327 | 29,999,673 |
| 3 | 2,734,841 | 27,265,159 | 481,724 | 29,518,276 | 927 | 29,999,073 | 444 | 29,999,556 |
| 4 | 3,457,975 | 26,542,025 | 612,747 | 29,387,253 | 994 | 29,999,006 | 475 | 29,999,525 |
| 6 | 5,320,713 | 24,679,287 | 991,606 | 29,008,394 | 1,097 | 29,998,903 | 529 | 29,999,471 |
| 7 | 6,261,628 | 23,738,372 | 1,226,695 | 28,773,305 | 1,136 | 29,998,864 | 546 | 29,999,454 |
| 9 | 7,916,882 | 22,083,118 | 1,740,067 | 28,259,933 | 1,221 | 29,998,779 | 587 | 29,999,413 |
| 10 | 8,658,021 | 21,341,979 | 2,009,835 | 27,990,165 | 1,274 | 29,998,726 | 612 | 29,999,388 |
| 12 | 10,131,849 | 19,868,151 | 2,591,299 | 27,408,701 | 1,701 | 29,998,299 | 710 | 29,999,290 |
| 13 | 10,917,472 | 19,082,528 | 2,923,699 | 27,076,301 | 2,146 | 29,997,854 | 796 | 29,999,204 |
| 15 | 12,646,165 | 17,353,835 | 3,730,089 | 26,269,911 | 3,921 | 29,996,079 | 1,153 | 29,998,847 |

**Table 8: Details of our third four datasets (set_9, set_10, set_11, and set_12). We report the accepted and the rejected pairs for edit distance thresholds of *E*=0 up to *E*=25 edits.**

| Dataset | Set_9 | | Set_10 | | Set_11 | | Set_12 | |
|---|---|---|---|---|---|---|---|---|
| *E* | Accepted | Rejected | Accepted | Rejected | Accepted | Rejected | Accepted | Rejected |
| 0 | 707,517 | 29,292,483 | 43,565 | 29,956,435 | 4,389 | 29,995,611 | 49 | 29,999,951 |
| 2 | 1,462,242 | 28,537,758 | 88,141 | 29,911,859 | 8,970 | 29,991,030 | 163 | 29,999,837 |
| 5 | 1,973,835 | 28,026,165 | 119,100 | 29,880,900 | 12,420 | 29,987,580 | 301 | 29,999,699 |
| 7 | 2,361,418 | 27,638,582 | 145,290 | 29,854,710 | 15,405 | 29,984,595 | 375 | 29,999,625 |
| 10 | 3,183,271 | 26,816,729 | 205,536 | 29,794,464 | 22,014 | 29,977,986 | 472 | 29,999,528 |
| 12 | 3,862,776 | 26,137,224 | 257,360 | 29,742,640 | 27,817 | 29,972,183 | 520 | 29,999,480 |
| 15 | 4,915,346 | 25,084,654 | 346,809 | 29,653,191 | 37,710 | 29,962,290 | 575 | 29,999,425 |
| 17 | 5,550,869 | 24,449,131 | 409,978 | 29,590,022 | 44,225 | 29,955,775 | 623 | 29,999,377 |
| 20 | 6,404,832 | 23,595,168 | 507,177 | 29,492,823 | 54,650 | 29,945,350 | 718 | 29,999,282 |
| 22 | 6,959,616 | 23,040,384 | 572,769 | 29,427,231 | 62,255 | 29,937,745 | 842 | 29,999,158 |
| 25 | 7,857,750 | 22,142,250 | 673,254 | 29,326,746 | 74,761 | 29,925,239 | 1,133 | 29,998,867 |

## 10 Evaluating the Number of Falsely-Accepted Sequence Pairs and Falsely-Rejected Sequence Pairs

We evaluate the number of falsely-accepted pairs and falsely-rejected pairs for Shouji, MAGNET, SHD (Xin et al., 2015), and GateKeeper (Alser et al., 2017). We list the number of falsely-accepted and falsely-rejected sequences in Table 9, Table 10, and Table 11 for read lengths of 100 bp, 150 bp, and 250 bp, respectively.

The false reject rate is the ratio of the number of similar sequences that are rejected (falsely-rejected pairs) by the filter and the number of similar sequences that are accepted by the optimal sequence alignment algorithm. The false reject rate should always be equal to 0%. Using our 12 low-edit and high-edit datasets for three different sequence lengths, we observe that Shouji, SHD, and GateKeeper do *not* filter out correct sequence pairs; hence, they provide a 0% false reject rate. The reason is the way we find the common subsequences. We always look for the subsequences that have the largest number of zeros, such that we maximize the number of matches and minimize the number of edits that cause the division of one long common sequence into shorter subsequences. However, this is not the case for MAGNET. We observe that MAGNET provides a very low false reject rate of less than 0.00045% for an edit distance threshold of at least 4% of the sequence length. This is due in large part to the greedy choice of always selecting the longest common subsequence regardless of its contribution to the total number of edits. On the contrary, Shouji always examines whether or not the selected 4-bit segment that has the largest number of zeros decreases the number of edits in the Shouji bit-vector before considering the 4-bit segment to be part of the common subsequences. In Fig. 7, we show an example of where MAGNET falsely considers two given sequences as dissimilar ones, while they differ by less than the edit distance threshold. This example shows that MAGNET's greedy approach of finding the common subsequences fails in finding the two common subsequences that are highlighted in blue. Instead, MAGNET finds another four shorter subsequences that result in increasing the number of mismatches in the MAGNET bit-vector.

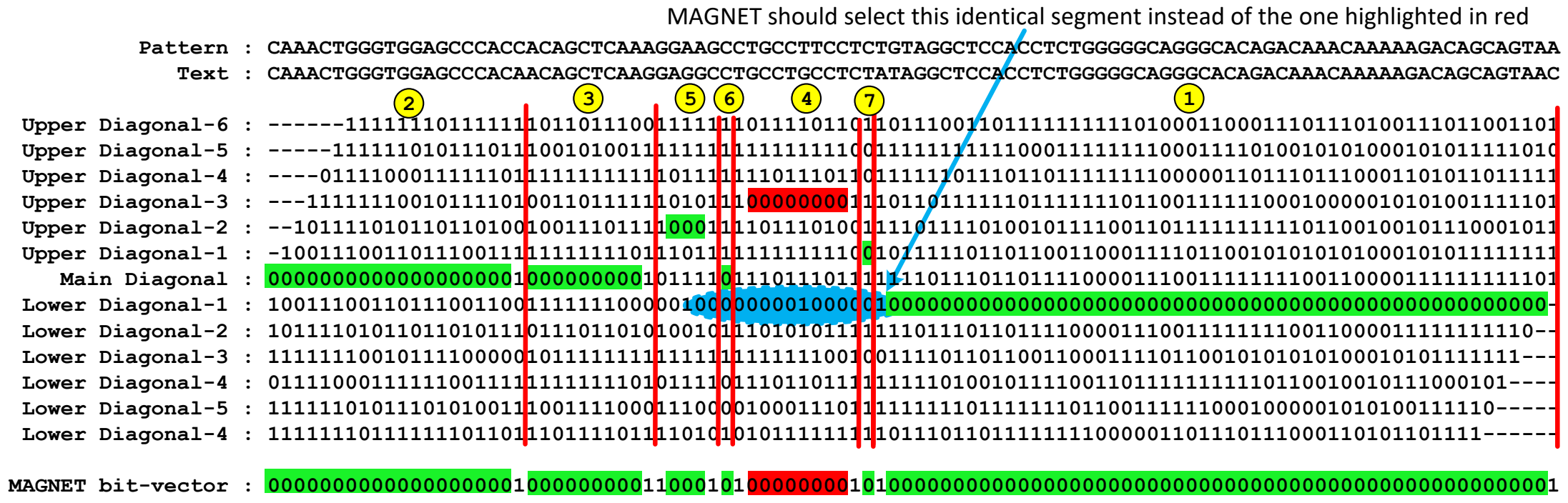

**Fig. 7: An example of a falsely-rejected sequence pair using the MAGNET algorithm for an edit distance threshold of 6. The random zeros (highlighted in red) confuse the MAGNET filter, causing it to select shorter segments of random zeros instead of a longer common subsequences (highlighted in blue).**

**Table 9: Details of evaluating the number of falsely-accepted sequence pairs (FA) and falsely-rejected sequence pairs (FR) of Shouji, MAGNET, GateKeeper, and SHD using four datasets, set_1, set_2, set_3, and set_4, with a read length of 100 bp.**

| | | Read Aligner | | Pre-alignment Filter | | | | | | | |
|---|---|---|---|---|---|---|---|---|---|---|---|
| | E | Edlib | | SHD | | GateKeeper | | MAGNET | | Shouji | |
| | | Accepted | Rejected | FA | FR | FA | FR | FA | FR | FA | FR |
| Set_1 | 0 | 381,901 | 29,618,099 | 10 | 0 | 0 | 0 | 963,941 | 0 | 0 | 0 |
| | 1 | 1,345,842 | 28,654,158 | 783,185 | 0 | 783,185 | 0 | 800,099 | 0 | 333,320 | 0 |
| | 2 | 3,266,455 | 26,733,545 | 2,704,128 | 0 | 2,704,128 | 0 | 1,876,518 | 0 | 1,283,004 | 0 |
| | 3 | 5,595,596 | 24,404,404 | 5,237,529 | 0 | 5,237,529 | 0 | 2,428,301 | 0 | 2,674,876 | 0 |
| | 4 | 7,825,272 | 22,174,728 | 8,231,507 | 0 | 8,231,507 | 0 | 2,662,902 | 1 | 4,399,886 | 0 |
| | 5 | 9,821,308 | 20,178,692 | 11,195,124 | 0 | 11,195,124 | 0 | 2,916,838 | 0 | 6,452,280 | 0 |
| | 6 | 11,650,490 | 18,349,510 | 13,781,651 | 0 | 13,781,651 | 0 | 3,406,303 | 4 | 9,373,309 | 0 |
| | 7 | 13,407,801 | 16,592,199 | 14,283,519 | 0 | 14,283,519 | 0 | 4,026,433 | 19 | 11,113,616 | 0 |
| | 8 | 15,152,501 | 14,847,499 | 13,814,295 | 0 | 13,814,295 | 0 | 4,745,672 | 27 | 11,990,529 | 0 |
| | 9 | 16,894,680 | 13,105,320 | 13,105,305 | 0 | 13,105,305 | 0 | 5,319,627 | 41 | 11,693,396 | 0 |
| | 10 | 18,610,897 | 11,389,103 | 11,389,103 | 0 | 11,389,103 | 0 | 5,673,172 | 31 | 10,664,722 | 0 |
| | E | Accepted | Rejected | FA | FR | FA | FR | FA | FR | FA | FR |
| Set_2 | 0 | 124,531 | 29,875,469 | 2 | 0 | 0 | 0 | 317,396 | 0 | 0 | 0 |
| | 1 | 441,927 | 29,558,073 | 276,271 | 0 | 276,271 | 0 | 265,663 | 0 | 114,225 | 0 |
| | 2 | 1,073,808 | 28,926,192 | 1,273,787 | 0 | 1,273,787 | 0 | 779,683 | 0 | 524,886 | 0 |
| | 3 | 2,053,181 | 27,946,819 | 3,370,661 | 0 | 3,370,661 | 0 | 1,257,472 | 0 | 1,494,883 | 0 |
| | 4 | 3,235,057 | 26,764,943 | 6,695,487 | 0 | 6,695,487 | 0 | 1,621,885 | 1 | 3,085,801 | 0 |
| | 5 | 4,481,341 | 25,518,659 | 10,798,431 | 0 | 10,798,431 | 0 | 1,995,105 | 0 | 5,410,196 | 0 |
| | 6 | 5,756,432 | 24,243,568 | 15,305,752 | 0 | 15,305,752 | 0 | 2,574,171 | 2 | 9,218,900 | 0 |
| | 7 | 7,091,373 | 22,908,627 | 17,347,813 | 0 | 17,347,813 | 0 | 3,391,117 | 5 | 12,401,268 | 0 |
| | 8 | 8,531,811 | 21,468,189 | 18,015,876 | 0 | 18,015,876 | 0 | 4,485,756 | 19 | 14,865,877 | 0 |
| | 9 | 10,102,726 | 19,897,274 | 19,897,204 | 0 | 19,897,204 | 0 | 5,639,763 | 38 | 15,670,345 | 0 |
| | 10 | 11,807,488 | 18,192,512 | 18,192,512 | 0 | 18,192,512 | 0 | 6,691,920 | 52 | 15,222,777 | 0 |
| | E | Accepted | Rejected | FA | FR | FA | FR | FA | FR | FA | FR |
| Set_3 | 0 | 11,989 | 29,988,011 | 1 | 0 | 0 | 0 | 32,576 | 0 | 0 | 0 |
| | 1 | 44,565 | 29,955,435 | 30,065 | 0 | 30,065 | 0 | 27,639 | 0 | 13,060 | 0 |
| | 2 | 108,979 | 29,891,021 | 153,613 | 0 | 153,613 | 0 | 77,792 | 0 | 61,519 | 0 |
| | 3 | 206,903 | 29,793,097 | 466,411 | 0 | 466,411 | 0 | 133,654 | 0 | 200,269 | 0 |
| | 4 | 334,712 | 29,665,288 | 1,254,259 | 0 | 1,254,259 | 0 | 193,569 | 0 | 521,359 | 0 |
| | 5 | 490,670 | 29,509,330 | 2,767,674 | 0 | 2,767,674 | 0 | 268,750 | 0 | 1,206,373 | 0 |
| | 6 | 675,357 | 29,324,643 | 6,227,154 | 0 | 6,227,154 | 0 | 385,154 | 0 | 2,983,331 | 0 |
| | 7 | 891,447 | 29,108,553 | 9,695,580 | 0 | 9,695,580 | 0 | 585,853 | 0 | 5,431,357 | 0 |
| | 8 | 1,151,447 | 28,848,553 | 12,921,874 | 0 | 12,921,874 | 0 | 931,084 | 1 | 8,532,786 | 0 |
| | 9 | 1,469,996 | 28,530,004 | 28,529,540 | 0 | 28,529,540 | 0 | 1,466,018 | 9 | 11,228,839 | 0 |
| | 10 | 1,868,827 | 28,131,173 | 28,131,173 | 0 | 28,131,173 | 0 | 2,251,403 | 6 | 13,630,704 | 0 |
| | E | Accepted | Rejected | FA | FR | FA | FR | FA | FR | FA | FR |
| Set_4 | 0 | 11 | 29,999,989 | 0 | 0 | 0 | 0 | 7 | 0 | 0 | 0 |
| | 1 | 18 | 29,999,982 | 14 | 0 | 14 | 0 | 5 | 0 | 2 | 0 |
| | 2 | 24 | 29,999,976 | 155 | 0 | 155 | 0 | 2 | 0 | 15 | 0 |
| | 3 | 27 | 29,999,973 | 1,196 | 0 | 1,196 | 0 | 4 | 0 | 216 | 0 |
| | 4 | 29 | 29,999,971 | 7,436 | 0 | 7,436 | 0 | 13 | 0 | 1,986 | 0 |
| | 5 | 34 | 29,999,966 | 32,792 | 0 | 32,792 | 0 | 82 | 0 | 10,551 | 0 |
| | 6 | 83 | 29,999,917 | 155,134 | 0 | 155,134 | 0 | 298 | 0 | 57,258 | 0 |
| | 7 | 177 | 29,999,823 | 417,444 | 0 | 417,444 | 0 | 1,030 | 0 | 214,005 | 0 |
| | 8 | 333 | 29,999,667 | 1,031,480 | 0 | 1,031,480 | 0 | 3,129 | 0 | 675,029 | 0 |
| | 9 | 711 | 29,999,289 | 29,997,022 | 0 | 29,997,022 | 0 | 8,234 | 0 | 1,742,476 | 0 |
| | 10 | 1,627 | 29,998,373 | 29,998,373 | 0 | 29,998,373 | 0 | 19,013 | 0 | 3,902,535 | 0 |

**Table 10: Details of evaluating the number of falsely-accepted sequence pairs (FA) and falsely-rejected sequence pairs (FR) of Shouji, MAGNET, GateKeeper, and SHD using four datasets, set_5, set_6, set_7, and set_8, with a read length of 150 bp.**

| | E | Read Aligner | | Pre-alignment Filter | | | | | | | |
|---|---|---|---|---|---|---|---|---|---|---|---|
| | | Edlib | | SHD | | GateKeeper | | MAGNET | | Shouji | |
| | | Accepted | Rejected | FA | FR | FA | FR | FA | FR | FA | FR |
| Set_5 | 0 | 1,440,497 | 28,559,503 | 0 | 0 | 0 | 0 | 428,412 | 0 | 0 | 0 |
| | 1 | 1,868,909 | 28,131,091 | 173,573 | 0 | 173,573 | 0 | 156,891 | 0 | 113,519 | 0 |
| | 3 | 2,734,841 | 27,265,159 | 2,080,279 | 0 | 2,080,279 | 0 | 725,873 | 0 | 1,539,365 | 0 |
| | 4 | 3,457,975 | 26,542,025 | 4,023,762 | 0 | 4,023,762 | 0 | 1,064,344 | 0 | 3,042,831 | 0 |
| | 6 | 5,320,713 | 24,679,287 | 9,258,602 | 0 | 9,258,602 | 0 | 1,430,272 | 0 | 6,025,592 | 0 |
| | 7 | 6,261,628 | 23,738,372 | 12,481,853 | 0 | 12,481,853 | 0 | 1,532,024 | 2 | 8,219,336 | 0 |
| | 9 | 7,916,882 | 22,083,118 | 22,076,837 | 0 | 22,076,837 | 0 | 1,874,734 | 20 | 14,568,337 | 0 |
| | 10 | 8,658,021 | 21,341,979 | 21,341,979 | 0 | 21,341,979 | 0 | 2,194,275 | 10 | 16,920,389 | 0 |
| | 12 | 10,131,849 | 19,868,151 | 19,868,151 | 0 | 19,868,151 | 0 | 3,294,672 | 42 | 18,270,597 | 0 |
| | 13 | 10,917,472 | 19,082,528 | 19,082,528 | 0 | 19,082,528 | 0 | 4,066,617 | 46 | 18,095,207 | 0 |
| | 15 | 12,646,165 | 17,353,835 | 17,353,835 | 0 | 17,353,835 | 0 | 5,810,797 | 62 | 16,993,568 | 0 |
| | E | Accepted | Rejected | FA | FR | FA | FR | FA | FR | FA | FR |
| Set_6 | 0 | 248,920 | 29,751,080 | 0 | 0 | 0 | 0 | 75,136 | 0 | 0 | 0 |
| | 1 | 324,056 | 29,675,944 | 31,406 | 0 | 31,406 | 0 | 28,456 | 0 | 20,294 | 0 |
| | 3 | 481,724 | 29,518,276 | 440,577 | 0 | 440,577 | 0 | 131,460 | 0 | 309,015 | 0 |
| | 4 | 612,747 | 29,387,253 | 1,023,901 | 0 | 1,023,901 | 0 | 199,248 | 0 | 718,847 | 0 |
| | 6 | 991,606 | 29,008,394 | 4,165,422 | 0 | 4,165,422 | 0 | 334,729 | 0 | 2,222,934 | 0 |
| | 7 | 1,226,695 | 28,773,305 | 7,137,889 | 0 | 7,137,889 | 0 | 405,052 | 0 | 3,762,706 | 0 |
| | 9 | 1,740,067 | 28,259,933 | 28,215,257 | 0 | 28,215,257 | 0 | 600,124 | 0 | 10,299,935 | 0 |
| | 10 | 2,009,835 | 27,990,165 | 27,990,165 | 0 | 27,990,165 | 0 | 753,866 | 2 | 13,826,393 | 0 |
| | 12 | 2,591,299 | 27,408,701 | 27,408,701 | 0 | 27,408,701 | 0 | 1,336,246 | 10 | 17,542,652 | 0 |
| | 13 | 2,923,699 | 27,076,301 | 27,076,301 | 0 | 27,076,301 | 0 | 1,835,774 | 19 | 18,371,563 | 0 |
| | 15 | 3,730,089 | 26,269,911 | 26,269,911 | 0 | 26,269,911 | 0 | 3,354,276 | 33 | 19,528,254 | 0 |
| | E | Accepted | Rejected | FA | FR | FA | FR | FA | FR | FA | FR |
| Set_7 | 0 | 444 | 29,999,556 | 0 | 0 | 0 | 0 | 251 | 0 | 0 | 0 |
| | 1 | 695 | 29,999,305 | 104 | 0 | 104 | 0 | 77 | 0 | 94 | 0 |
| | 3 | 927 | 29,999,073 | 191 | 0 | 191 | 0 | 68 | 0 | 180 | 0 |
| | 4 | 994 | 29,999,006 | 643 | 0 | 643 | 0 | 53 | 0 | 421 | 0 |
| | 6 | 1,097 | 29,998,903 | 47,924 | 0 | 47,924 | 0 | 57 | 0 | 19,097 | 0 |
| | 7 | 1,136 | 29,998,864 | 175,481 | 0 | 175,481 | 0 | 74 | 0 | 70,540 | 0 |
| | 9 | 1,221 | 29,998,779 | 29,595,345 | 0 | 29,595,345 | 0 | 461 | 0 | 857,547 | 0 |
| | 10 | 1,274 | 29,998,726 | 29,998,726 | 0 | 29,998,726 | 0 | 1,017 | 0 | 1,829,338 | 0 |
| | 12 | 1,701 | 29,998,299 | 29,998,299 | 0 | 29,998,299 | 0 | 4,218 | 0 | 4,893,299 | 0 |
| | 13 | 2,146 | 29,997,854 | 29,997,854 | 0 | 29,997,854 | 0 | 8,620 | 0 | 6,955,205 | 0 |
| | 15 | 3,921 | 29,996,079 | 29,996,079 | 0 | 29,996,079 | 0 | 31,783 | 0 | 12,854,488 | 0 |
| | E | Accepted | Rejected | FA | FR | FA | FR | FA | FR | FA | FR |
| Set_8 | 0 | 201 | 29,999,799 | 0 | 0 | 0 | 0 | 126 | 0 | 0 | 0 |
| | 1 | 327 | 29,999,673 | 58 | 0 | 58 | 0 | 42 | 0 | 43 | 0 |
| | 3 | 444 | 29,999,556 | 90 | 0 | 90 | 0 | 35 | 0 | 83 | 0 |
| | 4 | 475 | 29,999,525 | 267 | 0 | 267 | 0 | 28 | 0 | 137 | 0 |
| | 6 | 529 | 29,999,471 | 18,110 | 0 | 18,110 | 0 | 25 | 0 | 6,259 | 0 |
| | 7 | 546 | 29,999,454 | 79,418 | 0 | 79,418 | 0 | 27 | 0 | 27,092 | 0 |
| | 9 | 587 | 29,999,413 | 29,698,666 | 0 | 29,698,666 | 0 | 108 | 0 | 404,742 | 0 |
| | 10 | 612 | 29,999,388 | 29,999,388 | 0 | 29,999,388 | 0 | 231 | 0 | 935,486 | 0 |
| | 12 | 710 | 29,999,290 | 29,999,290 | 0 | 29,999,290 | 0 | 965 | 0 | 2,514,950 | 0 |
| | 13 | 796 | 29,999,204 | 29,999,204 | 0 | 29,999,204 | 0 | 2,018 | 0 | 3,693,298 | 0 |
| | 15 | 1,153 | 29,998,847 | 29,998,847 | 0 | 29,998,847 | 0 | 8,448 | 0 | 8,034,737 | 0 |

**Table 11: Details of evaluating the number of falsely-accepted sequence pairs (FA) and falsely-rejected sequence pairs (FR) of Shouji, MAGNET, GateKeeper, and SHD using four datasets, set_9, set_10, set_11, and set_12, with a read length of 250 bp.**

| | E | Read Aligner | | Pre-alignment Filter | | | | | | | |
|---|---|---|---|---|---|---|---|---|---|---|---|
| | | Edlib | | SHD | | GateKeeper | | MAGNET | | Shouji | |
| | | Accepted | Rejected | FA | FR | FA | FR | FA | FR | FA | FR |
| Set_9 | 0 | 707,517 | 29,292,483 | 0 | 0 | 0 | 0 | 479,104 | 0 | 0 | 0 |
| | 2 | 1,462,242 | 28,537,758 | 238,368 | 0 | 238,368 | 0 | 143,066 | 0 | 174,366 | 0 |
| | 5 | 1,973,835 | 28,026,165 | 1,546,126 | 0 | 1,546,126 | 0 | 226,864 | 0 | 1,071,218 | 0 |
| | 7 | 2,361,418 | 27,638,582 | 3,933,916 | 0 | 3,933,916 | 0 | 347,819 | 1 | 2,775,419 | 0 |
| | 10 | 3,183,271 | 26,816,729 | 26,816,729 | 0 | 26,816,729 | 0 | 624,927 | 1 | 6,669,084 | 0 |
| | 12 | 3,862,776 | 26,137,224 | 26,137,224 | 0 | 26,137,224 | 0 | 825,468 | 9 | 11,147,373 | 0 |
| | 15 | 4,915,346 | 25,084,654 | 25,084,654 | 0 | 25,084,654 | 0 | 1,066,633 | 14 | 18,406,823 | 0 |
| | 17 | 5,550,869 | 24,449,131 | 24,449,131 | 0 | 24,449,131 | 0 | 1,235,999 | 23 | 20,971,826 | 0 |
| | 20 | 6,404,832 | 23,595,168 | 23,595,168 | 0 | 23,595,168 | 0 | 1,695,351 | 35 | 22,223,170 | 0 |
| | 22 | 6,959,616 | 23,040,384 | 23,040,384 | 0 | 23,040,384 | 0 | 2,241,984 | 42 | 22,271,215 | 0 |
| | 25 | 7,857,750 | 22,142,250 | 22,142,250 | 0 | 22,142,250 | 0 | 3,514,515 | 54 | 21,849,454 | 0 |
| | E | Accepted | Rejected | FA | FR | FA | FR | FA | FR | FA | FR |
| Set_10 | 0 | 43,565 | 29,956,435 | 0 | 0 | 0 | 0 | 28,540 | 0 | 0 | 0 |
| | 2 | 88,141 | 29,911,859 | 13,092 | 0 | 13,092 | 0 | 8,367 | 0 | 11,238 | 0 |
| | 5 | 119,100 | 29,880,900 | 113,106 | 0 | 113,106 | 0 | 14,685 | 0 | 77,095 | 0 |
| | 7 | 145,290 | 29,854,710 | 364,611 | 0 | 364,611 | 0 | 24,919 | 0 | 227,073 | 0 |
| | 10 | 205,536 | 29,794,464 | 29,794,464 | 0 | 29,794,464 | 0 | 45,768 | 0 | 782,844 | 0 |
| | 12 | 257,360 | 29,742,640 | 29,742,640 | 0 | 29,742,640 | 0 | 63,557 | 2 | 2,195,021 | 0 |
| | 15 | 346,809 | 29,653,191 | 29,653,191 | 0 | 29,653,191 | 0 | 92,443 | 1 | 7,573,911 | 0 |
| | 17 | 409,978 | 29,590,022 | 29,590,022 | 0 | 29,590,022 | 0 | 116,740 | 1 | 11,603,069 | 0 |
| | 20 | 507,177 | 29,492,823 | 29,492,823 | 0 | 29,492,823 | 0 | 165,502 | 2 | 16,075,487 | 0 |
| | 22 | 572,769 | 29,427,231 | 29,427,231 | 0 | 29,427,231 | 0 | 217,274 | 6 | 19,167,498 | 0 |
| | 25 | 673,254 | 29,326,746 | 29,326,746 | 0 | 29,326,746 | 0 | 376,323 | 7 | 24,778,497 | 0 |
| | E | Accepted | Rejected | FA | FR | FA | FR | FA | FR | FA | FR |
| Set_11 | 0 | 4,389 | 29,995,611 | 0 | 0 | 0 | 0 | 2,933 | 0 | 0 | 0 |
| | 2 | 8,970 | 29,991,030 | 1,405 | 0 | 1,405 | 0 | 890 | 0 | 1,173 | 0 |
| | 5 | 12,420 | 29,987,580 | 12,185 | 0 | 12,185 | 0 | 1,704 | 0 | 8,489 | 0 |
| | 7 | 15,405 | 29,984,595 | 41,555 | 0 | 41,555 | 0 | 2,644 | 0 | 24,946 | 0 |
| | 10 | 22,014 | 29,977,986 | 29,977,986 | 0 | 29,977,986 | 0 | 4,759 | 0 | 145,053 | 0 |
| | 12 | 27,817 | 29,972,183 | 29,972,183 | 0 | 29,972,183 | 0 | 6,729 | 1 | 833,703 | 0 |
| | 15 | 37,710 | 29,962,290 | 29,962,290 | 0 | 29,962,290 | 0 | 9,498 | 0 | 5,088,387 | 0 |
| | 17 | 44,225 | 29,955,775 | 29,955,775 | 0 | 29,955,775 | 0 | 12,134 | 0 | 9,832,285 | 0 |
| | 20 | 54,650 | 29,945,350 | 29,945,350 | 0 | 29,945,350 | 0 | 18,366 | 0 | 16,815,067 | 0 |
| | 22 | 62,255 | 29,937,745 | 29,937,745 | 0 | 29,937,745 | 0 | 25,411 | 2 | 20,798,178 | 0 |
| | 25 | 74,761 | 29,925,239 | 29,925,239 | 0 | 29,925,239 | 0 | 44,377 | 1 | 26,094,659 | 0 |
| | E | Accepted | Rejected | FA | FR | FA | FR | FA | FR | FA | FR |
| Set_12 | 0 | 49 | 29,999,951 | 0 | 0 | 0 | 0 | 53 | 0 | 0 | 0 |
| | 2 | 163 | 29,999,837 | 71 | 0 | 71 | 0 | 44 | 0 | 55 | 0 |
| | 5 | 301 | 29,999,699 | 249 | 0 | 249 | 0 | 49 | 0 | 161 | 0 |
| | 7 | 375 | 29,999,625 | 698 | 0 | 698 | 0 | 48 | 0 | 212 | 0 |
| | 10 | 472 | 29,999,528 | 29,999,528 | 0 | 29,999,528 | 0 | 42 | 0 | 5,627 | 0 |
| | 12 | 520 | 29,999,480 | 29,999,480 | 0 | 29,999,480 | 0 | 45 | 0 | 64,225 | 0 |
| | 15 | 575 | 29,999,425 | 29,999,425 | 0 | 29,999,425 | 0 | 82 | 0 | 775,314 | 0 |
| | 17 | 623 | 29,999,377 | 29,999,377 | 0 | 29,999,377 | 0 | 175 | 0 | 2,052,498 | 0 |
| | 20 | 718 | 29,999,282 | 29,999,282 | 0 | 29,999,282 | 0 | 417 | 0 | 5,679,869 | 0 |
| | 22 | 842 | 29,999,158 | 29,999,158 | 0 | 29,999,158 | 0 | 593 | 0 | 10,277,297 | 0 |
| | 25 | 1,133 | 29,998,867 | 29,998,867 | 0 | 29,998,867 | 0 | 1,174 | 0 | 19,676,652 | 0 |

## 11 Evaluating the Number of Falsely-Accepted and Falsely-Rejected Pairs Using Single End and Paired End Reads

We assess the accuracy of Shouji using both single end and paired end reads. We first map 3' reads from ERR240727.fastq (i.e., reads from ERR240727_2.fastq) to the human reference genome (GRCh37) using mrFAST (Alkan et al., 2009) with an edit distance threshold of 2. We then use the first 30 million read-reference pairs that are produced by mrFAST before performing alignment to examine the filtering accuracy of Shouji. In Table 12, we show the number of falsely-accepted and falsely-rejected pairs of Shouji using these 30 million pairs over different edit distance thresholds. Generating the read-reference pairs in this way allows us to examine the filtering accuracy of Shouji using both aligned (i.e., pairs that have edits no more than the allowed edit distance threshold) and unaligned (i.e., pairs that have edits more than the allowed edit distance threshold) pairs. We use the same method to generate set_1 from ERR240727_1.fastq, as we describe in Section 3.1 in the main manuscript. We observe that the accuracy of Shouji using 3' reads from ERR240727.fastq remains almost the same as that of Shouji when we use 5' reads from ERR240727.fastq (which we show in Table 9 when we use set_1). Next, we map both 5' reads and 3' reads from ERR240727.fastq to the human reference genome using the mrFAST mapper in paired end mode. We then use the first 30 million read-reference pairs that are produced by mrFAST before performing alignment to examine the filtering accuracy of Shouji. In Table 13, we show the number of falsely-accepted and falsely-rejected pairs of Shouji using these 30 million pairs. We observe the results are similar when using paired end reads as when using single end reads. Based on Table 12 and Table 13, we conclude that the evaluation of our pre-alignment filter does not depend on the paired end sequencing or paired end reads. Similarly with any dynamic programming sequence alignment algorithm, Shouji always examines a single reference segment with a single read individually and independently from the way this pair is generated. The read mapper is responsible for generating the read-reference pairs that must be verified using a dynamic programming sequence alignment algorithm. Shouji examines these pairs (before using the computationally-expensive sequence alignment algorithms) regardless of the algorithm (e.g., single end read mapping or paired end read mapping) used to generate these pairs.

**Table 12: Number of falsely-accepted and falsely-rejected sequence pairs of Shouji using single end reads from ERR240727_2.fastq mapped to the human reference genome. We use Edlib (Šošić and Šikić, 2017) to generate the ground truth edit distance value for each sequence pair.**

| *E* | Edlib baseline | | Shouji | | | |
|---|---|---|---|---|---|---|
| | Aligned | Unaligned | Aligned | Unaligned | Falsely-Accepted | Falsely-Rejected |
| 0 | 206,252 | 29,793,748 | 206,252 | 29,793,748 | 0 | 0 |
| 1 | 1,359,165 | 28,640,835 | 1,680,722 | 28,319,278 | 321,557 | 0 |
| 2 | 3,308,445 | 26,691,555 | 4,562,146 | 25,437,854 | 1,253,701 | 0 |
| 3 | 5,673,028 | 24,326,972 | 8,290,885 | 21,709,115 | 2,617,857 | 0 |
| 4 | 7,929,996 | 22,070,004 | 12,171,061 | 17,828,939 | 4,241,065 | 0 |
| 5 | 9,920,919 | 20,079,081 | 16,051,171 | 13,948,829 | 6,130,252 | 0 |
| 6 | 11,710,868 | 18,289,132 | 20,532,091 | 9,467,909 | 8,821,223 | 0 |
| 7 | 13,409,936 | 16,590,064 | 23,845,857 | 6,154,143 | 10,435,921 | 0 |
| 8 | 15,078,030 | 14,921,970 | 26,405,117 | 3,594,883 | 11,327,087 | 0 |
| 9 | 16,727,424 | 13,272,576 | 27,901,872 | 2,098,128 | 11,174,448 | 0 |
| 10 | 18,339,408 | 11,660,592 | 28,680,484 | 1,319,516 | 10,341,076 | 0 |

**Table 13: Number of falsely-accepted and falsely-rejected sequence pairs of Shouji using paired end reads from ERR240727.fastq mapped to the human reference genome. We use Edlib (Šošić and Šikić, 2017) to generate the ground truth edit distance value for each sequence pair.**

| *E* | Edlib baseline | | Shouji | | | |
|---|---|---|---|---|---|---|
| | Aligned | Unaligned | Aligned | Unaligned | Falsely-Accepted | Falsely-Rejected |
| 0 | 0 | 30,000,000 | 0 | 30,000,000 | 0 | 0 |
| 1 | 373,921 | 29,626,079 | 453,808 | 29,546,192 | 79,887 | 0 |
| 2 | 1,318,319 | 28,681,681 | 1,947,127 | 28,052,873 | 628,808 | 0 |
| 3 | 3,207,952 | 26,792,048 | 5,224,261 | 24,775,739 | 2,016,309 | 0 |
| 4 | 5,500,950 | 24,499,050 | 9,227,434 | 20,772,566 | 3,726,484 | 0 |
| 5 | 7,709,237 | 22,290,763 | 13,305,866 | 16,694,134 | 5,596,629 | 0 |
| 6 | 9,698,512 | 20,301,488 | 18,208,145 | 11,791,855 | 8,509,633 | 0 |
| 7 | 11,529,693 | 18,470,307 | 22,281,600 | 7,718,400 | 10,751,907 | 0 |
| 8 | 13,293,029 | 16,706,971 | 25,736,052 | 4,263,948 | 12,443,023 | 0 |
| 9 | 15,041,936 | 14,958,064 | 27,833,759 | 2,166,241 | 12,791,823 | 0 |
| 10 | 16,782,466 | 13,217,534 | 28,890,050 | 1,109,950 | 12,107,584 | 0 |

## 12 FPGA Acceleration of Shouji and MAGNET

We analyze the benefits of accelerating the CPU implementation of our pre-alignment filters Shouji and MAGNET using FPGA hardware. As we show in Table 14, our hardware accelerators are two to three orders of magnitude faster than the equivalent CPU implementations of Shouji and MAGNET.

**Table 14: Execution time (in seconds) of the CPU implementations of Shouji and MAGNET filters and that of their hardware-accelerated versions (using a single filtering unit).**

| *E* | Shouji-CPU | Shouji-FPGA | Speedup | MAGNET-CPU | MAGNET-FPGA | Speedup |
|---|---|---|---|---|---|---|
| | | | *Sequence Length = 100* | | | |
| 2 | 474.27 | 2.89 | 164.11x | 632.02 | 2.89 | 218.69x |
| 5 | 1,305.15 | 2.89 | 451.61x | 1,641.57 | 2.89 | 568.02x |
| | | | *Sequence Length = 250* | | | |
| 2 | 1,689.09 | 2.89* | 584.46x | 5,567.62 | 2.89* | 1,926.51x |
| 5 | 6,096.61 | 2.89* | 2,109.55x | 14,328.28 | 2.89* | 4,957.88x |

* Estimated based on the resource utilization and data throughput

## 13 Execution time breakdown of Read Mapping combined with Shouji

We provide the total runtime breakdown of mrFAST (v. 2.6.1) (Alkan et al., 2009) and BWA-MEM (Li, 2013) with Shouji as a pre-alignment filter. We break down the execution time of read mapping with Shouji into 1) read-reference pair generation time, 2) Shouji filtering time, 3) Shouji pre-processing time, 4) Shouji transfer time, and 5) dynamic programming alignment time. The sum of these five runtime values provides the total execution time of read mapping with Shouji as a pre-alignment filter (8$^{th}$ column of Table 15 entitled total execution time). We provide the total execution time breakdown of mrFAST (v. 2.6.1 that includes FastHASH (Xin et al., 2013)) (Alkan et al., 2009) and BWA-MEM (Li, 2013) with Shouji compared to the baseline (i.e., the last column of Table 15 represents the runtime of mrFAST and BWA-MEM without Shouji) in Table 15. We map all reads from ERR240727_1 (100 bp) to GRCh37 with an edit distance threshold of 2% and 5%. Based on Table 15, we make the following key observation: the dynamic programming alignment time drops by a factor of 4-24 (the 7$^{th}$ column of Table 15 compared with the 10$^{th}$ column of Table 15) after integrating Shouji with read mapping as a pre-alignment step.

We conclude that the ability of Shouji to accelerate read mapping scales very well over a wide range of edit distance threshold values.

**Table 15: Total execution time breakdown (in seconds) of mrFAST and BWA-MEM with and without Shouji, for an edit distance threshold of 2% and 5%. The green shaded columns represent the processing time spent by each step of the original read mapper (without Shouji). The orange and blue shaded columns represent the processing time spent by each step of the accelerated read mapper (with the addition of Shouji as a pre-alignment step). The orange shaded columns represent the processing time spent by Shouji on the FPGA board and the host CPU.**

| | *E* | Read mapping time with Shouji | | | | | | Read mapping time without Shouji (baseline) | | |
|---|---|---|---|---|---|---|---|---|---|---|
| | | Read-ref pair generation time | Shouji (FPGA) filtering time | Shouji (CPU) pre-processing | Shouji (CPU) Transfer time | Alignment time | Total execution time | Read-ref pair generation time | Alignment time | Total execution time |
| mrFAST | 2 | 175.02 | 0.0616 | 3.2239 | 0.2919 | 16.6929 | 195.2902 | 175.02 | 67.08 | 242.1 |
| mrFAST | 5 | 198.02 | 1.3176 | 53.9911 | 6.2457 | 242.8571 | 502.4315 | 198.02 | 2333.99 | 2532.01 |
| BWA-MEM | 2 | 622.1 | 0.0010 | 0.0516 | 0.0050 | 4.8219 | 626.9794 | 622.1 | 46.02 | 668.12 |
| BWA-MEM | 2* | 623.03 | 0.0124 | 0.6477 | 0.0622 | 2.0729 | 625.8252 | 623.03 | 47.08 | 670.11 |
| BWA-MEM | 5 | 649.02 | 0.0010 | 0.0521 | 0.0050 | 4.7089 | 653.7870 | 649.02 | 46.12 | 695.14 |
| BWA-MEM | 5* | 650.01 | 0.0129 | 0.6740 | 0.0647 | 1.9190 | 652.6806 | 650.01 | 46.08 | 696.09 |

## 14 Edlib, Parasail, SHD, mrFAST, and BWA-MEM Configurations

In Table 16, we list the software packages that we cover in our performance evaluation, including their version numbers and function calls used.

**Table 16: Read aligners and pre-alignment filters used in our performance evaluations.**

**Edlib: November 5 2017**

Banded Levenshtein Distance:

```
EdlibAlignResult resultEdlib = edlibAlign(RefSeq, ReadLength, ReadSeq, ReadLength, edlibNewAlignConfig(ErrorThreshold,
EDLIB_MODE_NW, EDLIB_TASK_PATH, NULL, 0));
edlibFreeAlignResult(resultEdlib);
if (resultEdlib.editDistance!= -1)
            Accepted =1;
else      Accepted =0;
```

Banded Levenshtein Distance with backtracking:

```
EdlibAlignResult resultEdlib = edlibAlign(RefSeq, ReadLength, ReadSeq, ReadLength, edlibNewAlignConfig(ErrorThreshold,
EDLIB_MODE_NW, EDLIB_TASK_PATH, NULL, 0));
char* cigar = edlibAlignmentToCigar(resultEdlib.alignment, resultEdlib.alignmentLength, EDLIB_CIGAR_STANDARD);
free(cigar);
edlibFreeAlignResult(resultEdlib);
```

**Parasail: January 7 2018**

```
function = parasail_lookup_function("nw_banded");
result = function(RefSeq, ReadLength,   ReadSeq, ReadLength,10, 1, ErrorThreshold,&parasail_blosum62);
if(parasail_result_is_trace(result)==1){
    parasail_traceback_generic(RefSeq, ReadLength, ReadSeq, ReadLength, "Query:", "Target:", &parasail_blosum62, result, '|', ':', '.', 50, 14, 0);
    if (result->score != 0) {
        cigar2=parasail_result_get_cigar(result, RefSeq, ReadLength, ReadSeq, ReadLength, &parasail_blosum62);
        parasail_cigar_free(cigar2);
    }
}
```

**SHD: November 7 2017, compiled using g++-4.9**

```
for (k=1;k<=1+ (ReadLength/128);k++)
    totalEdits= totalEdits + (bit_vec_filter_sse1(read_t, ref_t, length, ErrorThreshold));
```

**mrFAST: November 29 2017**

```
./mrfast-2.6.1.0/mrfast --search human_g1k_v37.fasta --seq ../ERR240727_1_100bp.fastq -e 2
```

The human reference genome can be downloaded from:
ftp://ftp.ncbi.nlm.nih.gov/1000genomes/ftp/technical/reference/human_g1k_v37.fasta.gz

Extracting read-reference pairs:

1- Add the following to line 1786 of https://github.com/BilkentCompGen/mrfast/blob/master/MrFAST.c
2- Extract reference segment:
   `for (n = 0; n < 100; n++) printf("%d", _msf_refGen[n + genLoc + _msf_refGenOffset - 1 - leftSeqLength]);`
3- Extract read sequence:
   `printf("\t%s\n", _tmpSeq);`

**BWA-MEM: November 25 2018**

```
./bwa mem -w 3 ../human_g1k_v37.fasta ../../../Desktop/Filters_29_11_2016/ERR240727_1_100bp.fastq
```

Report all secondary alignments:

```
./bwa mem -a -w 3 ../human_g1k_v37.fasta ../../../Desktop/Filters_29_11_2016/ERR240727_1_100bp.fastq
```

Extracting read-reference pairs:

1- Add the following code between line 166 and line 167 of https://github.com/lh3/bwa/blob/master/bwa.c
2- Extract reference segment:
   `for (i = 0; i < rlen; ++i) putchar("ACGTN"[(int)rseq[i]]); putchar('\t');`
3- Extract read sequence:
   `for (i = 0; i < l_query; ++i) putchar("ACGTN"[(int)query[i]]); putchar('\n');`

## REFERENCES

Alkan, C., Kidd, J. M., Marques-Bonet, T., Aksay, G., Antonacci, F., Hormozdiari, F., Kitzman, J. O., Baker, C., Malig, M. and Mutlu, O. (2009) Personalized copy number and segmental duplication maps using next-generation sequencing, *Nature genetics*, **41**, 1061-1067.

Alser, M., Hassan, H., Xin, H., Ergin, O., Mutlu, O. and Alkan, C. (2017) GateKeeper: a new hardware architecture for accelerating pre-alignment in DNA short read mapping, *Bioinformatics*, **33**, 3355-3363.

Alser, M., Mutlu, O. and Alkan, C. (July 2017) Magnet: Understanding and improving the accuracy of genome pre-alignment filtering, *Transactions on Internet Research* **13**.

Li, H. (2013) Aligning sequence reads, clone sequences and assembly contigs with BWA-MEM, *arXiv preprint arXiv:1303.3997*.

McNamara, M. (2001) IEEE Standard Verilog Hardware Description Language. The Institute of Electrical and Electronics Engineers, *Inc. IEEE Std*, 1364-2001.

Šošić, M. and Šikić, M. (2017) Edlib: a C/C++ library for fast, exact sequence alignment using edit distance, *Bioinformatics*, **33**, 1394-1395.

Xilinx (November 17, 2014) 7 Series FPGAs Configurable Logic Block User Guide. Xilinx.

Xin, H., Greth, J., Emmons, J., Pekhimenko, G., Kingsford, C., Alkan, C. and Mutlu, O. (2015) Shifted Hamming Distance: A Fast and Accurate SIMD-Friendly Filter to Accelerate Alignment Verification in Read Mapping, *Bioinformatics*, **31**, 1553-1560.

Xin, H., Lee, D., Hormozdiari, F., Yedkar, S., Mutlu, O. and Alkan, C. (2013) Accelerating read mapping with FastHASH, *BMC genomics*, **14**, S13.